\documentclass[conference]{IEEEtran}
\IEEEoverridecommandlockouts
\usepackage{cite}
\usepackage{amsmath,amssymb,amsfonts}
\usepackage{algorithmic}
\usepackage{graphicx}
\usepackage{textcomp}
\usepackage{xcolor}

\usepackage{epstopdf}
\usepackage{makecell}
\usepackage{amsmath}
\usepackage{bm}
\usepackage{algorithmic}
\usepackage[ruled,vlined]{algorithm2e}
\usepackage{subfigure}

\def\BibTeX{{\rm B\kern-.05em{\sc i\kern-.025em b}\kern-.08em
    T\kern-.1667em\lower.7ex\hbox{E}\kern-.125emX}}
\begin{document}

\title{cuFasterTucker: A Stochastic Optimization Strategy for Parallel Sparse FastTucker Decomposition on GPU Platform
}

\author{\IEEEauthorblockN{1\textsuperscript{st} Zixuan Li}
\IEEEauthorblockA{
\textit{Hunan University}\\
Changsha, Hunan, China \\
zixuanli@hnu.edu.cn}
\and
\IEEEauthorblockN{2\textsuperscript{nd} Mingxing Duan}
\IEEEauthorblockA{
	\textit{Hunan University}\\
	Changsha, Hunan, China \\
	duanmingxing16@nudt.edu.cn}
\and
\IEEEauthorblockN{3\textsuperscript{rd} Wangdong Yang}
\IEEEauthorblockA{
\textit{Hunan University}\\
Changsha, Hunan, China \\
yangwangdong@hnu.edu.cn}
\and
\IEEEauthorblockN{4\textsuperscript{th} Kenli Li}
\IEEEauthorblockA{
\textit{Hunan University}\\
Changsha, Hunan, China \\
lkl@hnu.edu.cn}
}

\maketitle

\begin{abstract}
Currently, the size of scientific data is growing at an unprecedented rate. 
Data in the form of tensors exhibit high-order, high-dimensional, and highly sparse features.
Although tensor-based analysis methods are very effective, 
the large increase in data size makes the original tensor impossible to process.
Tensor decomposition decomposes a tensor into multiple low-rank matrices or tensors that can be exploited by tensor-based analysis methods.
Tucker decomposition is such an algorithm, which decomposes a $n$-order tensor into $n$ low-rank factor matrices and a low-rank core tensor.
However, most Tucker decomposition methods are accompanied by huge intermediate variables and huge computational load, 
making them unable to process high-order and high-dimensional tensors.

In this paper, we propose FasterTucker decomposition based on FastTucker decomposition, 
which is a variant of Tucker decomposition.
And an efficient parallel FasterTucker decomposition algorithm cuFasterTucker on GPU platform is proposed. 
It has very low storage and computational requirements, and effectively solves the problem of high-order and high-dimensional sparse tensor decomposition.
Compared with the state-of-the-art algorithm, it achieves a speedup of around $15X$ and $7X$ in updating the factor matrices and updating the core matrices, respectively.
\end{abstract}

\begin{IEEEkeywords}
Parallel Computing, Tensor decomposition.
\end{IEEEkeywords}

\section{Introduction}
Tensors are extensions of vectors and matrices, and are a general term for three-order or higher-order data\cite{kolda2009tensor}.
Data derived from the real world is usually high-order and can naturally be represented by tensors\cite{ahmadi2021randomized}.
For example, an image is a $3$-order tensor, and a video is a $4$-order tensor.
Tensors can clearly represent the complex interaction of multiple features of an entity, which cannot be represented using linear or planar data forms.
As tensor data becomes higher in order and larger in scale, it is no longer practical to analyze full tensors directly.
The tensor decomposition, which uses multiple low-rank feature matrices or tensors to represent the original tensor and preserves the multi-order structural information of the original tensor, emerges as the times require.
At present, tensor methods based on tensor decomposition have been widely used in 
recommender system\cite{chen2021deep}, 
social network\cite{fernandes2021tensor},
psychological testing\cite{diez2018algebraic}, 
biology\cite{huang2021tensor}, 
stoichiometry\cite{phan2018error}, 
cryptography\cite{shin2022passwordtensor}, 
signal processing\cite{chen2021tensor}, 
deep learning \cite{yin2021towards,kaliyar2021deepfake}, 
numerical analysis \cite{tyrtyshnikov2020tensor} and other fields.

Tucker decomposition\cite{tucker1964extension} is one of the mainstream tensor decomposition algorithms, 
and it is a generalization of Singular Value Decomposition (SVD) for high-order data.
It decomposes an $n$-order tensor into $n$ low-rank factor matrices and a low-rank core tensor. 
These factor matrices extract important features of different orders respectively, 
and the core tensor reflects the interaction between different orders.
Tucker decomposition is usually implemented by the Higher Order Singular Value Decomposition (HOSVD) algorithm\cite{de2000multilinear} or the Higher Order Orthogonal Iteration (HOOI) algorithm\cite{de2000best}.
HOSVD algorithm flattens each order of the original tensor and does SVD operations, 
with huge memory overhead and computational complexity for intermediate matrices\cite{kang2012gigatensor}.
Some variants of HOSVD algorithm were proposed, such as 
Truncated HOSVD (T-HOSVD) \cite{haardt2008higher,balda2016first}, 
Sequentially Truncated HOSVD (ST-HOSVD)\cite{vannieuwenhoven2012new}, 
Hierarchical HOSVD\cite{grasedyck2010hierarchical}, 
but did not solve the problem fundamentally.
The operation of Tensor Multiplying Matrix Chain (TTMc) in HOOI algorithm also has huge intermediate variables and computational complexity, 
which is its computational bottleneck.
Some variants of the Alternating Least Squares(ALS)-based HOOI algorithm have attempted to address these issues, such as, P-Tucker\cite{oh2018scalable}, Vest\cite{park2021vest}, 
ParTi\cite{parti} and GTA\cite{oh2019high}, with little success.

The scale of data derived from multiple relational interactions is growing at an unprecedented rate, 
such as recommender systems\cite{ioannidis2019coupled}, 
Quality of Service (QoS)\cite{cheng2019personalized},
social networks~\cite{wang2019ho}, 
etc., forming High-Order, High-Dimension, and Sparse Tensor (HOHDST).
To compress and process these HOHDST, which cannot even be stored and computed in a single machine, 
we need efficient and scalable parallel algorithms.
Some parallel sparse Tucker decomposition algorithms on different parallel platforms have made some progress on the above problem.
But they still have high intermediate variable storage and high computational complexity and scalability issues.
The FastTucker proposed in \cite{li2022cu_fasttucker} replaces the $n$-order core tensor with $n$ low-rank core matrices, 
reduces the space complexity and computational complexity of Tucker decomposition from exponential to polynomial level, 
and keeps the solution space of Tucker decomposition unchanged.
However, its parallel algorithm cuFastTucker on the GPU platform has a large number of redundant calculations, 
and its utilization efficiency of the GPU is not ideal.

In this paper, we propose cuFasterTucker, a stochastic optimization strategy for parallel sparse FastTucker decomposition on GPU Platform.
It updates one factor matrix or core matrix at a time and fixes other factor matrices and core matrices, 
which ensures that it is a convex optimization problem every time.
cuFasterTucker uses the Balanced Compressed Sparse Fiber (B-CSF) tensor storage format\cite{nisa2019load}, 
which increases the memory access efficiency while ensuring a relatively balanced load, and reduces the calculation of shared intermediate variables.
More importantly, cufast stores reusable intermediate variables, which only occupy a small amount of memory, but greatly reduce its computational complexity.

Our main contributions are the following:

\begin{enumerate}
	
	\item \emph{Algorithm.} 
	We propose cuFasterTucker, a stochastic optimization strategy for parallel sparse FastTucker decomposition on GPU Platform.
	It reduces the computation of reusable intermediate variables and shared intermediate variables, and makes full use of GPU storage resources.
	It works fine on HOHDST data.
	
	\item \emph{Theory.}
	We analyze the computational complexity of the main process of cuFastTucker and cuFasterTucker, 
	and prove that cuFasterTucker reduces the computational complexity from $(N-1)|\Omega|\sum J_nR$ $to$ $\sum I_{n}J_{n}R$.
	
	\item \emph{Performance.} 
	Before cuFasterTucker, cuFastTucker was the known optimal parallel sparse Tucker decomposition algorithm capable of handling hosterge data.
	Experiments show that compared with cuFastTucker, 
	cuFasterTucker achieves about $15.0X$ and $7.0X$ speedup in update factor matrices and update core matrices, respectively.
	In addition, cuFasterTucker is more suitable for processing higher-order tensors than cuFastTucker, 
	because cuFasterTucker takes much less time than cuFastTucker.
	
\end{enumerate}

The code of cuFasterTucker used in this paper and a toy dataset are available at https://github.com/ZixuanLi-China/cuFasterTucker for reproducibility. 
The rest of this paper is organized as follows. 
Section \ref{Section Preliminaries} introduces the notations, definitions, tensor operations, and FastTucker decomposition.
Section \ref{Section Proposed Method} describes our proposed method FasterTucker decomposition. 
Section \ref{Section cuFasterTucker On GPU} describes our proposed fine-grained parallel sparse FasterTucker algorithm cuFasterTucker on GPU platform. 
Section \ref{Section Experiments} presents experimental results of cuFasterTucker andits contrasting algorithms.
And Section \ref{Section Conclusion} summarizes our work.

\section{Preliminaries}\label{Section Preliminaries}

We describe the notations in this paper in Section \ref{Section Notations},
definitions in Section \ref{Section Basic Definitions}, 
and FastTucker decomposition in \ref{Section FastTucker Decomposition}
and Stochastic Gradient Descent (SGD) based sparse FastTucker decomposition in \ref{Section Sparse FastTucker Decomposition}. 
Notations are summarized in Table \ref{Table Notations}.

\subsection{Notations}\label{Section Notations}

We denote tensors by bold Euler script letters (such as $\bm{\mathcal{X}}$), matrices by bold uppercase (such as $\textbf{A}$), vectors by bold lowercase (such as $\textbf{a}$), scalars by regular lowercase or uppercase (such as $N$ and $k$).
The elements in the tensor are denoted by the symbolic name of the tensor and the index, for example, $x_{i_{1},i_{2},\cdots,i_{n}}$ denotes the $(i_{1},i_{2},\cdots,i_{n})$th element of the tensor $\bm{\mathcal{X}}$.
And, $\textbf{a}_{i}$ denotes the $i$th row of matrix $\textbf{A}$,  $\textbf{a}_{:,j}$ denotes the $j$th column of matrix $\textbf{A}$.

\begin{table}[!htbp]
	\setlength{\abovedisplayskip}{0pt}
	\setlength{\belowdisplayskip}{0pt}
	\renewcommand{\arraystretch}{1.5}
	\caption{Table of symbols.}
	\centering
	\footnotesize
	\label{Table Notations}
	\tabcolsep1pt
	\begin{tabular}{cc}
		\hline
		\makecell[c]{Symbol}                         & \makecell[c]{Definition}\\
		\hline
		\makecell[c]{$\bm{\mathcal{X}}$}             & \makecell[c]{Input $N$-order tensor $\in$ $\mathbb{R}^{I_{1}\times I_{2}\times\cdots \times I_{N}}$;}\\
		\makecell[c]{$x_{i_{1},i_{2},\cdots,i_{n}}$} & \makecell[c]{$i_{1},i_{2},\cdots,i_{n}$th element of tensor $\mathcal{X}$}\\
		\makecell[c]{$\{N\}$}                        & \makecell[c]{The ordered set $\{1,2,\cdots,N-1,N\}$}\\
		\makecell[c]{$\{R\}$}                        & \makecell[c]{The core index set $\{1,2,\cdots,R-1,R\}$}\\
		\makecell[c]{$\{I_n\}$}                      & \makecell[c]{The index set $\{1,2,\cdots,I_n-1,I_n\}$}\\
		\makecell[c]{$\Omega$}                       & \makecell[c]{Index $(i_{1},\cdots,i_{n},\cdots,i_{N})$ of a tensor $\mathcal{X}$}\\	
		\makecell[c]{$\textbf{A}^{(n)}$}             & \makecell[c]{$n$th factor matrix $\in$ $\mathbb{R}^{I_{n}\times J_{n}}$}\\
		\makecell[c]{$\textbf{B}^{(n)}$}             & \makecell[c]{$n$th core matrix $\in$ $\mathbb{R}^{J_{n}\times R}$}\\		
		\makecell[c]{$a_{i_{n}, :}^{(n)}$}           & \makecell[c]{$i_{n}$th row vector $\in$ $\mathbb{R}^{J_{n}}$ of $\textbf{A}^{(n)}$}\\
		\makecell[c]{$b_{:,r}^{(n)}$}                & \makecell[c]{$r$th column vector $\in$ $\mathbb{R}^{J_{n}}$ of $\textbf{B}^{(n)}$}\\		
		\makecell[c]{$\circ$}                        & \makecell[c]{Outer product}\\		
		\makecell[c]{$\times$}                       & \makecell[c]{Matrix product}\\
		\makecell[c]{$\times_{(n)}$}                 & \makecell[c]{$n$-Mode Tensor-Matrix product}\\
		\makecell[c]{$\otimes$}                      & \makecell[c]{Kronecker product}\\
		\makecell[c]{$\%$}                           & \makecell[c]{Remainder}\\
		\hline
	\end{tabular}
\end{table}

\subsection{Basic Definitions}\label{Section Basic Definitions}
\newtheorem{definition}{Definition}
\begin{definition}[$n$-Mode Tensor Matricization]
	Given a $N$-order tensor $\bm{\mathcal{X}}$ $\in$ $\mathbb{R}^{I_{1}\times\cdots\times I_{N}}$,
	$n$-Mode Tensor Matricization refers to that $\bm{\mathcal{X}}$
	is unfolded in the $n$th order to form a low order matrix $\textbf{X}^{(n)}\in\mathbb{R}^{I_{n}\times I_{1}\cdots I_{n-1}\cdot I_{n+1}\cdot\cdots\cdot I_{N}}$, where $\textbf{X}^{(n)}$ stores all element of the $\bm{\mathcal{X}}\in\mathbb{R}$ and the matrix element $x^{(n)}_{i_{n},j}$ of $\textbf{X}^{(n)}$ at the position $j=1+\sum_{k=1,n\neq k}^{N}\left[(i_{k}-1)\mathop{\prod}_{m=1,m\neq n}^{k-1}I_{m}\right]$ contains the tensor element $x_{i_{1},i_{2},\cdots,i_{n},\cdots,i_{N}}$ of the tensor $\bm{\mathcal{X}}$.
	
\end{definition}

\begin{definition}[$n$-Mode Tensor Vectorization]
	Given a $N$-order tensor $\bm{\mathcal{X}}$ $\in$ $\mathbb{R}^{I_{1}\times\cdots\times I_{N}}$,
	$n$-Mode Tensor Vectorization refers to that $\bm{\mathcal{X}}$
	is unfolded in the $n$th order to form a vector $\textbf{x}^{(n)}$, where $\textbf{x}^{(n)}$ stores all element of the $\bm{\mathcal{X}}\in\mathbb{R}$ and 
	the vector element $x^{(n)}_{k}$ of $\textbf{x}^{(n)}$ at the position $k=(j-1)I_{n}+i$ contains the tensor element $\textbf{X}^{(n)}_{i,j}$
	of $n$th matricization $\textbf{X}^{(n)}$ of a tensor $\mathcal{X}$.
	
\end{definition}

\begin{definition}[$n$-Mode Tensor-Matrix product]
	Given a $N$-order tensor $\bm{\mathcal{X}}$ $\in$ $\mathbb{R}^{I_{1}\times\cdots\times I_{N}}$ and a matrix $\textbf{A}$ $\in$ $\mathbb{R}^{J_{n}\times I_{n}}$,
	$n$-Mode Tensor-Matrix product projects $\bm{\mathcal{X}}$ and $\textbf{A}$ to a new tensor $(\bm{\mathcal{X}}\times_{(n)} \textbf{A})$ $\in$ $\mathbb{R}^{I_{1}\times\cdots \times I_{n-1}\times J_{n}\times I_{n+1} \times\cdots  I_{N}}$ according to the coordinates, where $(\bm{\mathcal{X}}\times_{(n)} \textbf{A})_{i_{1}\times\cdots \times i_{n-1}\times j_{n}\times i_{n+1}\times\cdots\times  i_{N}}$ $=$ $\sum\limits_{i_{n}=1}^{I_{n}}$ $x_{i_{1}\times\cdots \times i_{n}\times \cdots\times i_{N}}$ $\cdot a_{j_{n},i_{n}}$.
	
\end{definition}

\begin{definition}[Rank-one Tensor]
	Given a $N$-order tensor $\bm{\mathcal{X}}$ $\in$ $\mathbb{R}^{I_{1}\times\cdots\times I_{N}}$,
	$\bm{\mathcal{X}}$ is Rank-one Tensor if it can be written as the outer product of N vectors, i.e., $\bm{\mathcal{X}}$ $=$ $b^{(1)} \circ b^{(2)} \circ \cdots \circ b^{(n)}$.
	
\end{definition}

\begin{definition}[The Rank of a Tensor]
	Given a $N$-order tensor $\bm{\mathcal{X}}$ $\in$ $\mathbb{R}^{I_{1}\times\cdots \times I_{N}}$,
	the rank of $\bm{\mathcal{X}}$, denoted $rank_{n}(\bm{\mathcal{X}})$, is defined as the smallest number of Rank-one Tensors that generate $\bm{\mathcal{X}}$ as their sum.
	
\end{definition}

\begin{definition}[$R$ Kruskal Product]
	Given $N$ matrices $\textbf{B}^{(n)}$ $\in$ $\mathbb{R}^{I_{n}\times R}$, $n \in \{N\}$,
	a $N$-order tensor $\widehat{\bm{\mathcal{X}}}$ $\in$ $\mathbb{R}^{I_{1}\times\cdots \times I_{N}}$ can be obtained by $R$ Kruskal Product: 
	$\widehat{\bm{\mathcal{X}}}=\sum_{r=1}^{R} b^{(1)}_{:,r}\circ\cdots\circ b^{(n)}_{:,r}\circ\cdots\circ b^{(N)}_{:,r}$.
\end{definition}

\begin{definition}[Khatri-Rao product]
	Given a matrice $\textbf{A}$ $\in$ $\mathbb{R}^{I\times K}$ and a matrice $\textbf{B}$ $\in$ $\mathbb{R}^{J\times K}$,
	Khatri-Rao product projects $\textbf{A}$ and $\textbf{B}$ to a new matrice $(\textbf{A} \odot \textbf{B})$
	$\in$ $\mathbb{R}^{IJ\times K}$ according to the coordinates, where $(\textbf{A} \odot \textbf{B})_{m,n}$ $=$ $\textbf{a}_{i/I,n}$ $\cdot$ $\textbf{b}_{m\%J,n}$.
\end{definition}

\begin{definition}[Kronecker product]
	Given a matrice $\textbf{A}$ $\in$ $\mathbb{R}^{I\times J}$ and a matrice $\textbf{B}$ $\in$ $\mathbb{R}^{K\times L}$,
	Kronecker product projects $\textbf{A}$ and $\textbf{B}$ to a new matrice $(\textbf{A} \otimes \textbf{B})$
 	$\in$ $\mathbb{R}^{IK\times JL}$ according to the coordinates, where $(\textbf{A} \otimes \textbf{B})_{m,n}$ $=$ $\textbf{a}_{m/I,n/J}$ $\cdot$ $\textbf{b}_{m\%K,n\%L}$.
\end{definition}

\begin{definition}[Tensor Approximation]
	Given a $N$-order sparse tensor $\bm{\mathcal{X}}$ $\in$ $\mathbb{R}^{I_{1}\times\cdots \times I_{N}}$,  the Tensor Approximation is to find a low-rank tensor $\widehat{\bm{\mathcal{X}}}$ $\in$ $\mathbb{R}^{I_{1}\times\cdots \times I_{N}}$ such that
	$\mathcal{D}(\bm{\mathcal{E}})$ small enough, where noisy tensor $\bm{\mathcal{E}}=$$\bm{\mathcal{X}}-\widehat{\bm{\mathcal{X}}}$$\in$ $\mathbb{R}^{I_{1} \times\cdots\times I_{N}}$ and $\mathcal{D}$ is a norm function.
	
\end{definition}

\subsection{FastTucker Decomposition}\label{Section FastTucker Decomposition}

Tensor decomposition is a solution to tensor approximation. 
Commonly used tensor decomposition methods include Canonical Polyadic (CP) decomposition, Tucker decomposition and FastTucker decomposition.
Our proposed method FasterTucker is based on FastTucker decomposition.
Given a $N$-order tensor $\bm{\mathcal{X}}$ $\in$ $\mathbb{R}^{I_{1}\times\cdots \times I_{N}}$, the goal of FastTucker decomposition is
to find $N$ factor matrices $\textbf{A}^{(n)}$ $\in$ $\mathbb{R}^{I_{n}\times J_{n}}$,$J_{n} \ll rank_{n}(\mathcal{X})$, $n \in \{N\}$ and $N$ core matrices $\textbf{B}^{(n)}$ $\in$ $\mathbb{R}^{J_{n}\times R}$, $n \in \{N\}$, such that:

\begin{equation}\label{fasttucker}
	\footnotesize
	\begin{aligned}
		\bm{\mathcal{X}}\approx\widehat{\bm{\mathcal{X}}}=&\bigg(\sum_{r=1}^{R} b^{(1)}_{:,r}\circ\cdots\circ b^{(n)}_{:,r}\circ\cdots\circ b^{(N)}_{:,r}\bigg) \times_{(1)}\textbf{A}^{(1)}\times_{(2)}\\
		&\cdots\times_{(n)}\textbf{A}^{(n)}\times_{(n+1)}\cdots\times_{(N)}\textbf{A}^{(N)}
	\end{aligned}
\end{equation}

In fact, FastTucker decomposition further decomposes the $N$-order core tensor in the tucker decomposition into N core matrices.

\begin{figure}[htbp]
	\centering
	\includegraphics[width=3.0in]{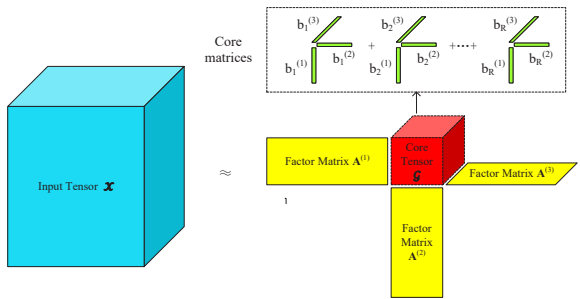}
	\caption{An example of FastTucker}
	\label{FastTucker_example}
\end{figure}

We let $\mathcal{D}$ be the Frobenius Norm, in order to make $\mathcal{D}(\bm{\mathcal{X}}-\widehat{\bm{\mathcal{X}})}$ as small as possible, it is transformed into the following optimization problem:

\begin{equation}\label{fasttucker_optimization}
	\footnotesize
	\begin{aligned}
		&\mathop{\arg\min}_{\textbf{A}^{(n)},n \in \{N\}, \textbf{B}^{(n)},n \in \{N\}}f\bigg(\bm{\mathcal{X}}, \big\{\textbf{A}^{(n)}\big\}, \big\{\textbf{B}^{(n)}\big\}, n \in\{N\} \bigg)\\
		&=\bigg\|\bm{\mathcal{X}}-\widehat{\bm{\mathcal{X}}}\bigg\|_{2}^{2}+\sum_{n=1}^{N}\lambda_{\textbf{A}^{(n)}}\|\textbf{A}^{(n)}\|_{2}^{2}+\sum_{n=1}^{N}\lambda_{\textbf{B}^{(n)}}\|\textbf{B}^{(n)}\|_{2}^{2}
	\end{aligned}
\end{equation}
where $\lambda_{\textbf{A}^{(n)}}, n \in \{N\}$ and $\lambda_{\textbf{B}^{(n)}}, n \in \{N\}$ are the regularization parameters for low-rank factor matrices $\textbf{A}^{(n)}, n \in \{N\}$ and core matrices $\textbf{B}^{(n)}, n \in \{N\}$, respectively.

The optimization objective (\ref{fasttucker_optimization}) involves variables multiplication,
which is non-convex.
The non-convex problem can be tackled by convex solution via updating a variable and fixing the others.
And the matricized version of equation (\ref{fasttucker}) is

\begin{equation}
	\footnotesize
	\begin{aligned}
		\widehat{\textbf{X}}^{(n)}=&\textbf{A}^{(n)}(\textbf{B}^{(n)}(\textbf{B}^{(N)}\odot\cdots \odot\textbf{B}^{(n+1)}\odot\textbf{B}^{(n-1)}\odot\cdots\\ &\odot\textbf{B}^{(1)})^{T})(\textbf{A}^{(N)}\otimes\cdots \otimes\textbf{A}^{(n+1)}\otimes\textbf{A}^{(n-1)}\otimes\cdots\\
		&\otimes\textbf{A}^{(1)})^{T}
	\end{aligned}
\end{equation}
where $\widehat{\textbf{X}}^{(n)}$ is $n$th matricization of tensor $\widehat{\bm{\mathcal{\bm{X}}}}$.
The optimization problem (\ref{fasttucker_optimization}) can be split into
updating low-rank factor matrices $\textbf{A}^{(n)}, n \in \{N\}$
and updating core matrices $\textbf{B}^{(n)}, n \in \{N\}$  as following:

\begin{equation}\label{low_rank_optimization}
	\footnotesize
	\begin{aligned}
		\mathop{\arg\min}_{\textbf{A}^{(n)},n \in\{N\}}
		&f\bigg(\textbf{A}^{(n)}\bigg|\textbf{X}^{(n)}, \big\{\textbf{A}^{(n)}\big\},\big\{\textbf{B}^{(n)}\big\}, n \in\{N\} \bigg)\\
		&=\bigg\|\textbf{X}^{(n)}- \widehat{\textbf{X}}^{(n)}\bigg\|_{2}^{2}+\lambda_{\textbf{A}^{(n)}}\|\textbf{A}^{(n)}\|_{2}^{2}
	\end{aligned}
\end{equation}

and
\begin{equation}\label{core_tensor_optimization}
	\footnotesize
	\begin{aligned}
		\mathop{\arg\min}_{\textbf{B}^{(n)},n \in\{N\}}
		&f\bigg(\textbf{B}^{(n)}\bigg|\textbf{X}^{(n)}, \big\{\textbf{A}^{(n)}\big\},\big\{\textbf{B}^{(n)}\big\}, n \in\{N\}\bigg)\\
		&=\bigg\|\textbf{X}^{(n)}- \widehat{\textbf{X}}^{(n)}\bigg\|_{2}^{2}+\lambda_{\textbf{B}^{(n)}}\|\textbf{B}^{(n)}\|_{2}^{2}
	\end{aligned}
\end{equation}
where $\widehat{\textbf{X}}^{(n)}$ $=$ $\textbf{A}^{(n)}\textbf{B}^{(n)}\textbf{Q}^{(n)^{T}}\textbf{S}^{(n)^{T}}$,
$\textbf{S}^{(n)}$ $=$ $\textbf{A}^{(N)}\otimes\cdots \otimes\textbf{A}^{(n+1)}\otimes\textbf{A}^{(n-1)}\otimes\cdots \otimes\textbf{A}^{(1)}$  and
$\textbf{Q}^{(n)}$ $=$ 
$\textbf{B}^{(N)}\odot\cdots \odot\textbf{B}^{(n+1)}\odot\textbf{B}^{(n-1)}\odot\cdots \odot\textbf{B}^{(1)}$.

For any $n \in \{N\}$, optimization objectives (\ref{low_rank_optimization}) and (\ref{core_tensor_optimization}) are both convex optimization problems, and the smaller the value of the loss function, the smaller the value of the loss function of optimization objective (\ref{fasttucker_optimization}).
Solving convex optimization problems can make the value of the loss function keep decreasing, and by solving these optimization objectives in a loop, the non-convex optimization objective can finally reach convergence.

\subsection{Sparse FastTucker Decomposition}\label{Section Sparse FastTucker Decomposition}

Note that the optimization objectives (\ref{fasttucker_optimization}), (\ref{low_rank_optimization}) and (\ref{core_tensor_optimization}) are calculated by all elements of $\bm{\mathcal{X}}$, and whole missing values of $\bm{\mathcal{X}}$ are regarded as zeros.
In fact, most tensors are extremely sparse, and it is unreasonable to treat all missing elements as zeros.
We denote the set of observable elements in $\bm{\mathcal{X}}$ as $\Omega$, and the number of elements in $\Omega$ as $|\Omega|$.
Sparse FastTucker decomposition finds $\textbf{A}^{(n)}, n \in \{N\}$ and $\textbf{B}^{(n)}, n \in \{N\}$ 
only by optimization objective (\ref{sparse_fasttucker_optimization}) with the elements present in $\bm{\mathcal{X}}$, and predicts values of missing elements after
$\textbf{A}^{(n)}, n \in \{N\}$ and $\textbf{B}^{(n)}, n \in \{N\}$ are found.

\begin{equation}\label{sparse_fasttucker_optimization}
	\footnotesize
	\begin{aligned}
		&\mathop{\arg\min}_{\textbf{A}^{(n)},n \in \{N\}, \textbf{B}^{(n)},n \in \{N\}}f\bigg(\bm{\mathcal{X}}, \big\{\textbf{A}^{(n)}\big\}, \big\{\textbf{B}^{(n)}\big\}, n \in\{N\} \bigg)\\
		=&\sum_{x_{i_1,\dots,i_N} \in \Omega }\bigg\|x_{i_1,\dots,i_N}- \widehat{x}_{i_1,\dots,i_N}\bigg\|_{2}^{2}+\sum_{n=1}^{N}\lambda_{\textbf{A}^{(n)}}\|\textbf{A}^{(n)}\|_{2}^{2}\\
		&+\sum_{n=1}^{N}\lambda_{\textbf{B}^{(n)}}\|\textbf{B}^{(n)}\|_{2}^{2}
	\end{aligned}
\end{equation}

In fact, the overall optimization problem (\ref{sparse_fasttucker_optimization}) can be decomposed into multiple single-element optimization problems. 
Similar to the dense FastTucker, the overall optimization problem (\ref{sparse_fasttucker_optimization}) can be solved by alternately solving convex factor matrix optimization problem (\ref{low_rank_optimization_single})
\begin{equation}\label{low_rank_optimization_single}
	\footnotesize
	\begin{aligned}
		&\mathop{\arg\min}_{\textbf{a}^{(n)}_{i_{n}},n \in\{N\}}
		f\bigg(\textbf{a}^{(n)}_{i_{n}}\bigg|x_{i_1,\dots,i_N}, \big\{\textbf{A}^{(n)}\big\}, \big\{\textbf{B}^{(n)}\big\}, n \in\{N\} \bigg)\\
		&=\bigg\|x_{i_1,\dots,i_N}- \widehat{x}_{i_1,\dots,i_N}\bigg\|_{2}^{2}
		+\lambda_{\textbf{a}^{(n)}_{i_{n}}}\|\textbf{a}^{(n)}_{i_{n}}\|_{2}^{2}
	\end{aligned}
\end{equation}
and convex core matrix optimization problem (\ref{core_tensor_optimization_single})
\begin{equation}\label{core_tensor_optimization_single}
	\footnotesize
	\begin{aligned}
		&\mathop{\arg\min}_{\textbf{b}^{(n)}_{:,r},n \in\{N\},r \in\{R\}}
		f\bigg(\textbf{b}^{(n)}_{:,r}\bigg|x_{i_1,\dots,i_N}, \big\{\textbf{A}^{(n)}\big\}, \big\{\textbf{B}^{(n)}\big\}, n \in\{N\} \bigg)\\
		&=\bigg\|x_{i_1,\dots,i_N}- \widehat{x}_{i_1,\dots,i_N}\bigg\|_{2}^{2}
		+\lambda_{\textbf{b}^{(n)}_{:,r}}\|\textbf{b}^{(n)}_{:,r}\|_{2}^{2}
	\end{aligned}
\end{equation}\\

In large-scale optimization scenarios,
SGD is a common strategy and promises to obtain the optimal accuracy via a certain number of training epoches.
An $M$ elements set $\Psi$ is randomly selected from the set $\Omega$, the learning rate is $\gamma$, and the SGD is presented as:
\begin{equation}
	\footnotesize
	\begin{aligned}\label{SGD}
		w&\leftarrow w-\gamma\frac{\partial f_{\Psi}(w)}{\partial w}\\
		&=w- \gamma\frac{1}{M}\sum_{i\in\Psi}\frac{\partial f_{i}(w)}{\partial w}.
	\end{aligned}
\end{equation}

The SGD for the approximated function $f\bigg(\textbf{a}^{(n)}_{i_{n}}\bigg|x_{i_1,\dots,i_N}, \big\{\textbf{a}^{(n)}_{i_{n}}\big\},\big\{\textbf{B}^{(n)}\big\}, n \in\{N\} \bigg)$ is deduced as:

\begin{equation}\label{Gradient_low_rank_cp}
	\footnotesize
	\begin{aligned}
		&\frac{\partial f\bigg(\textbf{a}^{(n)}_{i_{n}}\bigg|x_{i_1,\dots,i_N}, \big\{\textbf{a}^{(n)}_{i_{n}}\big\},\big\{\textbf{B}^{(n)}\big\}, n \in\{N\} \bigg)}{\partial \textbf{a}^{(n)}_{i_{n}}}\\
		=&\bigg(x_{i_1,\dots,i_N}-\textbf{a}^{(n)}_{i_{n}}\textbf{B}^{(n)}\textbf{Q}^{(n)^{T}}\textbf{s}^{(n)^{T}}_{i_1,\dots,i_N}\bigg)\\
		&\cdot\textbf{s}^{(n)}_{i_1,\dots,i_N}\textbf{Q}^{(n)}\textbf{B}^{(n)^{T}}+\lambda_{\textbf{a}^{(n)}_{i_{n}}}\textbf{a}^{(n)}_{i_{n}}\\
		=&-\bigg(x_{i_1,\dots,i_N}-\textbf{a}^{(n)}_{i_{n}}\big(\sum_{r=1}^{R}\textbf{b}^{(n)}_{:,r}\textbf{s}^{(n)}_{i_1,\dots,i_N}\textbf{q}^{(n)}_{:,r}\big)\bigg)\\
		&\cdot\big(\sum_{r=1}^{R}\textbf{b}^{(n)}_{:,r}\textbf{s}^{(n)}_{i_1,\dots,i_N}\textbf{q}^{(n)}_{:,r}\big)^{T}+\lambda_{\textbf{a}^{(n)}_{i_{n}}}\textbf{a}^{(n)}_{i_{n}}\\
	\end{aligned}
\end{equation}

and the SGD for the approximated function $f\bigg(\textbf{b}^{(n)}_{:,r}\bigg|x_{i_1,\dots,i_N}, 	\big\{\textbf{a}^{(n)}_{i_{n}}\big\},\big\{\textbf{b}^{(n)}_{:,r}\big\}\bigg)$ is deduced as:

\begin{equation}\label{Gradient_core_tensor_cp}
	\footnotesize
	\begin{aligned}
		&\frac{\partial f\bigg(\textbf{b}^{(n)}_{:,r}\bigg|x_{i_1,\dots,i_N}, \big\{\textbf{a}^{(n)}_{i_{n}}\big\},\big\{\textbf{b}^{(n)}_{:,r}\big\}\bigg)}{\partial \textbf{b}^{(n)}_{:,r}}\\
		=&-\bigg(x_{i_1,\dots,i_N}
		-\textbf{a}^{(n)}_{i_{n}}\textbf{B}^{(n)}\textbf{Q}^{(n)^{T}}\textbf{s}^{(n)^{T}}_{i_1,\dots,i_N}\bigg)\\
		&\cdot\textbf{a}^{(n)^{T}}_{(i_{n})}\textbf{s}^{(n)}_{i_1,\dots,i_N}\textbf{q}^{(n)}_{:,r}
		+\lambda_{\textbf{b}^{(n)}_{:,r}}\textbf{b}^{(n)}_{:,r}\\
		=&-\bigg(x_{i_1,\dots,i_N}
		-\textbf{a}^{(n)}_{i_{n}}\big(\sum_{r^{'}=1}^{R}\textbf{b}^{(n)}_{:,r^{'}}\textbf{s}^{(n)}_{i_1,\dots,i_N}\textbf{q}^{(n)}_{:,r^{'}}\big)\bigg)\\
		&\cdot\textbf{a}^{(n)^{T}}_{(i_{n})}\textbf{s}^{(n)}_{i_1,\dots,i_N}\textbf{q}^{(n)}_{:,r}
		+\lambda_{\textbf{b}^{(n)}_{:,r}}\textbf{b}^{(n)}_{:,r}\\
	\end{aligned}
\end{equation}

More importantly, in equations (\ref{Gradient_low_rank_cp}) and (\ref{Gradient_core_tensor_cp}), $\textbf{s}^{(n)}_{i_1,\dots,i_N}\textbf{q}^{(n)}_{:,r}$ has a simpler way of computing as shown in equation (\ref{s_q_simple}), which is also the core of FastTucker decomposition that reduces the Tucker decomposition from exponential computational complexity to polynomial computational complexity.
\begin{equation}\label{s_q_simple}
	\footnotesize
	\begin{aligned}
		\textbf{s}^{(n)}_{i_1,\dots,i_N}\textbf{q}^{(n)}_{:,r}=&
		(\textbf{a}^{(N)}_{i_N}\otimes\cdots \otimes\textbf{a}^{(n+1)}_{i_{n+1}}\otimes\textbf{a}^{(n-1)}_{i_{n-1}}\otimes\cdots\\ &\otimes\textbf{a}^{(1)}_{i_1})(\textbf{b}^{(N)}_{:,r}\otimes\cdots \otimes\textbf{b}^{(n+1)}_{:,r}\otimes\textbf{b}^{(n-1)}_{:,r}\\
		&\otimes\cdots \otimes\textbf{b}^{(1)}_{:,r})\\
		=&(\textbf{a}^{(N)}_{i_N}\textbf{b}^{(N)}_{:,r})\otimes \cdots 
		\otimes(\textbf{a}^{(n+1)}_{i_{n+1}}\textbf{b}^{(n+1)}_{:,r})\otimes\\
		&(\textbf{a}^{(n-1)}_{i_{n-1}}\textbf{b}^{(n-1)}_{:,r})
		\otimes \cdots \otimes (\textbf{a}^{(1)}_{i_1}\textbf{b}^{(1)}_{:,r})\\ 
		=&(\textbf{a}^{(N)}_{i_N}\textbf{b}^{(N)}_{:,r}) \cdots 
		(\textbf{a}^{(n+1)}_{i_{n+1}}\textbf{b}^{(n+1)}_{:,r})\cdot\\
		&(\textbf{a}^{(n-1)}_{i_{n-1}}\textbf{b}^{(n-1)}_{:,r})
		\cdots(\textbf{a}^{(1)}_{i_1}\textbf{b}^{(1)}_{:,r}) 
	\end{aligned}
\end{equation}

To minimize the optimization problem (\ref{sparse_fasttucker_optimization}), 
a SGD technique is used, which updates a factor matrix or a core matrix while keeping all others fixed.
Algorithm \ref{sgd_fasttucker} describes the conventional FastTucker decomposition algorithm. 
Algorithm \ref{sgd_fasttucker} is mainly divided into two parts, one is the update factor matrices module, the other is the update core matrices module.
In the update factor matrix module, Algorithm \ref{sgd_fasttucker} updates the factor matrix of each order in turn and fixes the factor matrix of other orders and all core matrices.
Next, traverse all element sets $\Psi_{i_{n}}^{(n)}$ and update the factor matrix.
It is worth noting that the index of the $n$th order of all elements $x_{i_1,\dots,i_N}$ contained in $\Psi_{i_{n}}^{(n)}$ is $i_{n}$, so that $\textbf{a}^{(n)}_{i_{n}}$ can be updated at the same time.
Similar to the update factor matrix module,
in the update core matrix module,
Algorithm \ref{sgd_fasttucker} updates the core matrices of each order in turn and fixes the core matrix of other orders and all factor matrices.
Next, traverse all element sets $\Psi$ and update the core matrix.
The difference is that the $x_{i_1,\dots,i_N}$ in $\Psi$ here can be completely randomly selected.
\begin{algorithm}[hptb]
	\caption{FastTucker Algorithm}
	\label{sgd_fasttucker}
	\footnotesize
	\vspace{.1cm}
	$\textbf{Input}$: Sparse tensor $\mathcal{X}$ $\in$ $\mathbb{R}^{I_{1}\times\cdots\times I_{N}}$, 
	initialized factor matrices $\textbf{A}^{(n)}$ $\in$ $\mathbb{R}^{I_{n}\times J_{n}}$, $n \in \{N\}$ and core matrices $\textbf{B}^{(n)}$ $\in$ $\mathbb{R}^{J_{n}\times R}$, $n \in \{N\}$,
	learning rates $\gamma_{\textbf{A}}$ and $\gamma_{\textbf{B}}$,
	regularization parameters $\lambda_{\textbf{A}}$ and $\lambda_{\textbf{A}}$.\\
	$\textbf{Output}$: Factor matrices $\textbf{A}^{(n)}$, $n$ $\in$ $\{N\}$ and core matrices $\textbf{B}^{(n)}$, $n$ $\in$ $\{N\}$.\\
	\begin{algorithmic}[1]
		\WHILE{$\mathcal{D}(\bm{\mathcal{E}})$ is not small enough}
		\FOR{$n$ from $1$ to $N$}
		\FOR{all sampling sets $\Psi_{i_{n}}^{(n)}$}
		\STATE Update $\textbf{a}^{(n)}_{i_{n}}$ according to equations (\ref{SGD}) \\
		and (\ref{Gradient_low_rank_cp}).
		\ENDFOR
		\ENDFOR
		\FOR{$n$ from $1$ to $N$}
		\FOR{all sampling sets $\Psi$}
		\STATE Update $\textbf{b}^{(n)}_{:,r}, r \in R$ according to equations \\
		(\ref{SGD}) and (\ref{Gradient_core_tensor_cp}).
		\ENDFOR
		\ENDFOR	
		\ENDWHILE	
	\end{algorithmic}
\end{algorithm}

\section{Proposed Method} \label{Section Proposed Method}

We describe FasterTucker, our proposed FastTucker decomposition based algorithm for sparse tensors.
FastTucker decomposition keeps the solution space of Tucker decomposition unchanged and reduces its exponential computational complexity to polynomial computational complexity, but there is still a lot of computational redundancy in FastTucker.
Our proposed FasterTucker avoids unnecessary computational redundancy in FastTucker.
We describe reusable intermediate variables and shared intermediate variables in Sections \ref{Section Reusable Intermediate Variables} and \ref{Section Invariant Intermediate Variables}, respectively. 
We present the fast decomposition algorithm in Section \ref{Section Sparse FasterTucker Decomposition} and perform a complexity analysis in Section \ref{Section Complexity Analysis}.

\subsection{Reusable Intermediate Variables} \label{Section Reusable Intermediate Variables}

It can be known from equations (\ref{Gradient_low_rank_cp}) and (\ref{Gradient_core_tensor_cp}) that the main calculation amount of the gradient is the calculation of $\textbf{s}^{(n)}_{i_1,\dots,i_N}\textbf{q}^{(n)}_{:,r}$.
At the same time according to equation (\ref{SGD}), the value of $\textbf{s}^{(n)}_{i_1,\dots,i_N}\textbf{q}^{(n)}_{:,r},n \in \{N\}, r \in \{R\}$ remains unchanged when $\textbf{a}^{(n)}_{i_n}, n \in \{N\}$ or $\textbf{b}^{(n)}_{:,r}, n \in \{N\}, r \in \{R\}$ is updated.
Further, it can be seen from equation (\ref{s_q_simple}) that $\textbf{s}^{(n)}_{i_1,\dots,i_N}\textbf{q}^{(n)}_{:,r}, n \in \{N\}, r \in \{R\}$ is formed by the permutation and combination of $\textbf{a}^{(n^)}_{i_{n}}\textbf{b}^{(n)}_{:,r}, n \in \{N\}, r \in \{R\}$.
Therefore, calculating each $\textbf{a}^{(n)}_{i_{n}}\textbf{b}^{(n)}_{:,r}, n \in \{N\}, r \in \{R\}$ in advance and then calling them when calculating $\textbf{s}^{(n)}_{i_1,\dots,i_N}\textbf{q}^{(n)}_{:,r}, n \in \{N\}, r \in \{R\}$, which can avoid a lot of repeated calculations.

\subsection{Shared Invariant Intermediate Variables} \label{Section Invariant Intermediate Variables}

Also note, for two non-zero elements $x_{i_1,\dots,i_{n_1},\dots,i_{N}}$ and $x_{i_1,\dots,i_{n_2},\dots,i_{N}}$ which update $\textbf{a}^{(n)}_{i_{n_1}}$ and $\textbf{a}^{(n)}_{i_{n_2}}$ respectively,
their intermediate matrix $\textbf{B}^{(n)}\textbf{Q}^{(n)^{T}}\textbf{s}^{(n)^{T}}_{i_1,\dots,i_N}$ is consistent and does not change after the update.
Therefore, updating $\textbf{a}^{(n)}_{i_{n_1}}$ and $\textbf{a}^{(n)}_{i_{n_2}}$ together can reduce the number of multiplication operations for $\textbf{B}^{(n)}\textbf{Q}^{(n)^{T}}\textbf{s}^{(n)^{T}}_{i_1,\dots,i_N}$.
So only need to calculate $\textbf{B}^{(n)}\textbf{Q}^{(n)^{T}}\textbf{s}^{(n)^{T}}_{i_1,\dots,i_N}$ once when updating $\textbf{a}^{(n)}_{i_{n^{'}}}, i_{n^{'}} \in \{I_n\}, x_{i_1,\dots,i_{n^{'}},\dots,i_{N}} \in \bm{\mathcal{X}}$, if put all $\{x_{i_1,\dots,i_{n^{'}},\dots,i_{N}}|i_{n^{'}} \in \{I_n\}, x_{i_1,\dots,i_{n^{'}},\dots,i_{N}} \in \bm{\mathcal{X}}\}$ into a $\Psi^{(n)}_{i_{n^{'}}}$.
When $\textbf{b}^{(n)}_{:,r}, n \in \{N\}, r \in \{R\}$ is updated, 
the elements set $\Psi^{(n)}_{i_{n^{'}}}$ is also used, 
then $\textbf{B}^{(n)}\textbf{Q}^{(n)^{T}}\textbf{s}^{(n)^{T}}_{i_1,\dots,i_N}$ can also be reused, 
and $\textbf{s}^{(n)}_{i_1,\dots,i_N}\textbf{q}^{(n)}_{:,r}, n \in \{N\}, r \in \{R\}$ can also be reused.

\subsection{Sparse FasterTucker Decomposition} \label{Section Sparse FasterTucker Decomposition}

On the basis of FastTucker, FasterTucker avoids the repeated calculation of the above two intermediate variables.
Algorithm \ref{sgd_fastertucker} describes the FasterTucker decomposition algorithm. 
Same as Algorithm \ref{sgd_fasttucker}, Algorithm \ref{sgd_fastertucker} is also divided into two parts: 
the factor matrices module and the core matrices module. 
But before that, 
Algorithm \ref{sgd_fastertucker} computes the reusable intermediate variables 
$\textbf{a}^{(n)}_{i_{n}}\textbf{b}^{(n)}_{:,r}, i_{n} \in \{I_{n}\}, n \in \{N\}, r \in \{R\}$ first. 
Algorithm \ref{sgd_fastertucker} updates each order in the same order and way as Algorithm \ref{sgd_fasttucker}, but the selection and processing of the element sets $\Psi^{(n)}_{i_{n^{'}}}$ is different.
Whether it is the factor matrices module or the kernel matrices module, the selection of the element set $\Psi^{(n)}_{i_{n^{'}}}$ is consistent. 
$\Psi^{(n)}_{i_{n^{'}}}$ fixes the indices of all order but the $n$th order, putting all elements $x_{i_1,\dots,i_{n^{'}},\dots,i_{N}}$ that meet that criteria into the same $\Psi^{(n)}_{i_{n^{'}}}$.
This allows all elements $x_{i_1,\dots,i_{n^{'}},\dots,i_{N}}$ in the elements set $\Psi^{(n)}_{i_{n^{'}}}$ to share $\textbf{B}^{(n)}\textbf{Q}^{(n)^{T}}\textbf{s}^{(n)^{T}}_{i_1,\dots,i_N}$,
avoiding the overhead of multiple computations.
For the factor matrices module, 
Algorithm \ref{sgd_fastertucker} does not update $\textbf{a}^{(n)}_{i_{n}}$ through the entire $\Psi_{i_{n}}^{(n)}$ as in Algorithm \ref{sgd_fasttucker}, but sequentially updates $\textbf{a}^{(n)}_{i_{n^{'}}}$ according to the $x_{i_1,\dots,i_{n^{'}},\dots,i_{N}}$ in $\Psi^{(n)}_{i_{n^{'}}}$.
For the factor matrices module, $\textbf{b}^{(n)}_{:,r}, r \in \{R\}$ can be updated by the whole $\Psi^{(n)}_{i_{n^{'}}}$.
And after updating the factor matrix or core matrix of $n$th order each time, the stored $\textbf{a}^{(n)}_{i_{n}}\textbf{b}^{(n)}_{:,r}, i_{n} \in \{I_{n}\}, r \in \{R\}$ needs to be updated.
\begin{algorithm}[hptb]
	\caption{FasterTucker Algorithm}
	\label{sgd_fastertucker}
	\footnotesize
	\vspace{.1cm}
	$\textbf{Input}$: Sparse tensor $\mathcal{X}$ $\in$ $\mathbb{R}^{I_{1}\times\cdots\times I_{N}}$, 
	initialized factor matrices $\textbf{A}^{(n)}$ $\in$ $\mathbb{R}^{I_{n}\times J_{n}}$, $n \in \{N\}$ and core matrices $\textbf{B}^{(n)}$ $\in$ $\mathbb{R}^{J_{n}\times R}$, $n \in \{N\}$,
	learning rates $\gamma_{\textbf{A}}$ and $\gamma_{\textbf{B}}$,
	regularization parameters $\lambda_{\textbf{A}}$ and $\lambda_{\textbf{A}}$.\\
	$\textbf{Output}$: Factor matrices $\textbf{A}^{(n)}$, $n$ $\in$ $\{N\}$ and core matrices $\textbf{B}^{(n)}$, $n$ $\in$ $\{N\}$.\\	
	\begin{algorithmic}[1]
		\FOR{$n$ from $1$ to $N$}
		\STATE Calculate and store $\textbf{a}^{(n)}_{i_{n}}\textbf{b}^{(n)}_{:,r}, i_{n} \in \{I_{n}\}, r \in \{R\}$.
		\ENDFOR
		\WHILE{$\mathcal{D}(\bm{\mathcal{E}})$ is not small enough}
		\FOR{$n$ from $1$ to $N$}
		\FOR{all sampling sets $\Psi^{(n)}_{i_{n^{'}}}$}
		\STATE Calculate $\textbf{s}^{(n)}_{i_1,\dots,i_N}\textbf{q}^{(n)}_{:,r}, r \in \{R\}$ by calling $\textbf{a}^{(n^{'})}_{i_{n^{'}}}\textbf{b}^{(n^{'})}_{:,r}, n \in \{N\}, n^{'}\neq n, r \in \{R\}$.
		\STATE Calculate $\textbf{B}^{(n)}\textbf{Q}^{(n)^{T}}\textbf{s}^{(n)^{T}}_{i_1,\dots,i_N}$.
		\FOR{all $x_{i_1,\dots,i_{n^{'}},\dots,i_{N}}$ in $\Psi^{(n)}_{i_{n^{'}}}$}
		\STATE Update $\textbf{a}^{(n)}_{i_{n^{'}}}$ according to equations (\ref{SGD}) \\
		and (\ref{Gradient_low_rank_cp}).
		\ENDFOR
		\ENDFOR
		\STATE Calculate and store $\textbf{a}^{(n)}_{i_{n}}\textbf{b}^{(n)}_{:,r}, i_{n} \in \{I_{n}\}, r \in \{R\}$.
		\ENDFOR
		\FOR{$n$ from $1$ to $N$}
		\FOR{all sampling sets $\Psi^{(n)}_{i_{n^{'}}}$}
		\STATE Calculate $\textbf{s}^{(n)}_{i_1,\dots,i_N}\textbf{q}^{(n)}_{:,r}, r \in \{R\}$ by calling $\textbf{a}^{(n^{'})}_{i_{n^{'}}}\textbf{b}^{(n^{'})}_{:,r}, n \in \{N\}, n^{'}\neq n, r \in \{R\}$.
		\STATE Calculate $\textbf{B}^{(n)}\textbf{Q}^{(n)^{T}}\textbf{s}^{(n)^{T}}_{i_1,\dots,i_N}$.
		\STATE Update $\textbf{b}^{(n)}_{:,r}, n \in \{N\}, r \in \{R\}$ according to equations
		(\ref{SGD})	and (\ref{Gradient_core_tensor_cp}).
		\ENDFOR
		\STATE Calculate and store $\textbf{a}^{(n)}_{i_{n}}\textbf{b}^{(n)}_{:,r}, i_{n} \in \{I_{n}\}, r \in \{R\}$.
		\ENDFOR	
		\ENDWHILE	
	\end{algorithmic}
\end{algorithm}

\subsection{Complexity Analysis} \label{Section Complexity Analysis}

In the basic FastTucker algorithm, the most important calculation amount in the process of updating $\textbf{a}^{(n)}_{i_n}, n \in \{N\}$ or $\textbf{b}^{(n)}_{:,r}, n \in \{N\}, r \in \{R\}$ is the calculation of $\textbf{a}^{(n)}_{i_{n}}\textbf{b}^{(n)}_{:,r}, n \in N, r \in R$.
The biggest difference between FastTucker algorithm and FasterTucker algorithm is the calculation method of $\textbf{a}^{(n)}_{i_{n}}\textbf{b}^{(n)}_{:,r}, n \in \{N\}, r \in \{R\}$.
The FastTucker algorithm calculates $\textbf{a}^{(n^{'})}_{i_{n^{'}}}\textbf{b}^{(n^{'})}_{:,r}, n^{'} \in \{N\}, n^{'} \neq n, r \in \{R\}$ as needed when updating $\textbf{a}^{(n)}_{i_n}$ with $x_{i_1,\dots,i_{n},\dots,i_{N}}$,
its multiplication calculation amount is $\sum\limits_{n^{'}\neq n} J_{n^{'}}R$.
For $N$ orders and $|\Omega|$ non-zero values in the sparse tensor $\bm{\mathcal{X}}$, the overall multiplication cost is $(N-1)|\Omega|\sum J_nR$.
And FasterTucker algorithm calculates the required $\textbf{a}^{(n^{'})}_{i_{n^{'}}}\textbf{b}^{(n^{'})}_{:,r}, n^{'} \in \{N\}, n^{'} \neq n, r \in \{R\}$ in advance and calls them when them are used,
the multiplication calculation of the whole process is $\sum I_{n}J_{n}R$.
Obviously, $\sum I_{n}J_{n}R$ $<$ $max(I_{n})\sum J_{n}R$ $<$ $(N-1)|\Omega|\sum J_nR$.
For shared intermediate variables, 
the multiplication computational cost of the $n$th order $\textbf{B}^{(n)}\textbf{Q}^{(n)^{T}}\textbf{s}^{(n)^{T}}_{i_1,\dots,i_N}$ is $J_nR+N-2$, and the multiplication computational cost of the entire FastTucker is $\sum_{n=1}^{N}\sum_{\Psi_{i_{n}}^{(n)}\in \Omega}|\Psi_{i_{n}}^{(n)}|(J_nR+N-2)$. 
However, the FasterTucker only needs to calculate $\textbf{B}^{(n)}\textbf{Q}^{(n)^{T}}\textbf{s}^{(n)^{T}}_{i_1,\dots,i_N}$ once for each $\Psi^{(n)}_{i_{n^{'}}}$, 
and the multiplication calculation amount of the shared intermediate variable of the FasterTucker is $\sum_{n=1}^{N}\sum_{\Psi^{(n)}_{i_{n^{'}}}\in \Omega}(J_nR+N-2)$.

\section{cuFasterTucker On GPU} \label{Section cuFasterTucker On GPU}
We describe our fine-grained parallel sparse FastTucker decomposition algorithm cuFasterTucker on GPU in detail.
We describe the tensor storage format used by cuFasterTucker in Section \ref{Section Tensor Storage Format}.
cuFasterTucker uses a two-level parallelism model: 
worker parallelization is described in Section \ref{Section Worker Parallelization}, 
and thread parallelization is described in Section \ref{Section Thread Parallelization}.
And we describe the GPU technology used by cuFasterTucker in Section \ref{Section Fine-grained Parallelization}.
Finally, we describe the specific implementation of cufaster with algorithms in Section \ref{Section Overview}.

\subsection{Tensor Storage Format} \label{Section Tensor Storage Format}

The elements set $\Psi^{(n)}_{i_{n^{'}}}$ used by FasterTucker is different from that of FastTucker. 
Except for the index of a certain order, 
all the other indexes of the elements $x_{i_1,\dots,i_{n^{'}},\dots,i_{N}}$ in $\Psi^{(n)}_{i_{n^{'}}}$ are fixed. 
In fact, such $x_{i_1,\dots,i_{n^{'}},\dots,i_{N}}$ is a subset of one of the slices of the tensor $\mathcal{X}$.
Considering the particularity of random sampling in FasterTucker algorithm, it is more appropriate to use CSF format to store sparse tensors.
At the same time, using CSF as a storage format offers opportunities to reduce operations and random memory accesses.
However, real-world tensors tend to follow a power-law distribution, which can lead to widespread load imbalances between threads or thread blocks on parallel platforms.
For example, the number of non-zero elements in slices allocated to different threads or thread blocks may vary widely.
B-CSF is used to solve such problems, 
it divides heavy slices into multiple sub-slices, 
heavy fibers into multiple sub-fibers, 
and heavy tensors into multiple sub-tensors.
This results in a relatively uniform distribution of non-zero elements among multiple threads or thread blocks.
The cuFastertucker of our proposed cuda platform uses the B-CSF tensor storage format. 
Although the division of sub-slices slightly increases the amount of computation, 
it is negligible compared to the benefits brought by load balancing.

\begin{figure}[htbp]
	\centering
	\subfigure[CSF]{
		\label{CSF}
		\includegraphics[width=1.5in]{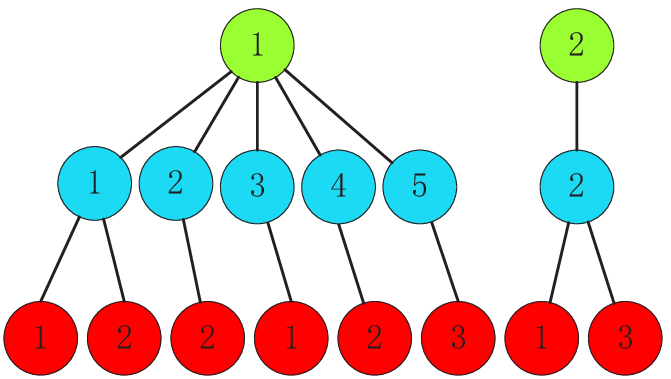}}
	~
	\subfigure[B-CSF]{
		\label{B-CSF}
		\includegraphics[width=1.5in]{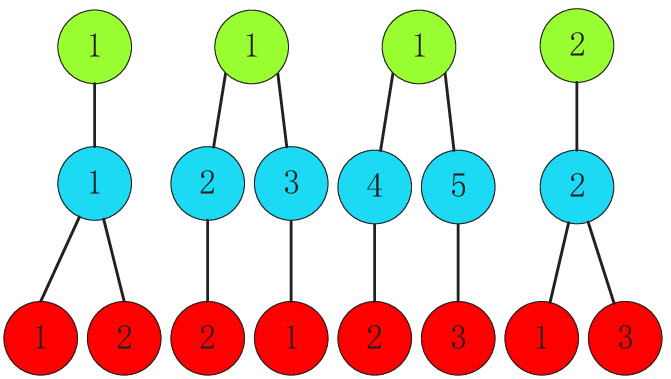}}
	\caption{Convergence curves of cuFasterTucker and its contrasting algorithm, they all set $J_n=32, n \in \{N\}$ and $R=32$.}
	\label{CSF and B-CSF}
\end{figure}

\subsection{Worker Parallelization} \label{Section Worker Parallelization}
GPU is a single-instruction multi-device architecture that can execute multiple blocks in parallel, 
and a block contains multiple threads. 
Multiple threads form a thread group, 
and the size of the scheduling unit (Warp) of the current mainstream Nvidia GPU architecture is $32$ threads. 
Therefore, it has better performance when the number of threads in a thread block is a divisor or multiple of $32$, 
and the number of threads in a block is a multiple of $32$. 
We divide each block into multiple thread groups and treat a thread group as a worker.
Each worker is responsible for a sub-tensor of tensor $\mathcal{X}$ at a time, 
and this sub-tensor is obtained by fixing a certain index of a certain dimension of tensor $\mathcal{X}$. 
We use the B-CSF format to store the tensor $\mathcal{X}$, 
which makes the number of non-zero elements contained in each sub-tensor relatively balanced. 
Therefore, the loads of multiple thread groups in multiple blocks are not much different, 
which avoids the problem of very unbalanced load among thread groups.

\subsection{Thread Parallelization} \label{Section Thread Parallelization}
Although each worker is responsible for a sub-tensor, 
instead of processing multiple non-zero elements at the same time, 
all non-zero elements of the sub-tensor are sequentially traversed.
If a single threads group updates multiple non-zero elements at the same time, 
it will bring new load balancing problems, 
multiple calculations of intermediate variables, 
and additional communication overhead.
We make each thread in the thread group process a scalar, 
then a threads group can process a vector.
We set all $J_n, n \in \{N\}$ to be divisors of $32$ or multiples of $32$.
If $J_n, n \in \{N\}$ is a divisor of $32$ including $32$, set the size of the thread group to $J_n, n \in \{N\}$. 
A thread group processes exactly one vector of length $J_n, n \in \{N\}$, the size of $\textbf{a}^{(n)}_{i_{n}}, n \in \{N\}$ or $\textbf{b}^{(n)}_{:,r}, n \in \{N\}, r \in \{R\}$, at a time. 
If $J_n, n \in \{N\}$ is a multiple of $32$ excluding $32$, set the thread group size to $32$. 
A thread group processes a vector of length $32$ at a time, 
the thread group processes multiple times to complete the processing of a vector of length $J_n, n \in \{N\}$.

\subsection{Fine-grained Parallelization} \label{Section Fine-grained Parallelization}
According to the algorithmic characteristics of the Fastertucker and the architectural characteristics of the GPU, 
we perform fine-grained parallel optimization on the proposed cuFasterTucker.
The major optimization techniques in cuFasterTucker are concluded as:
\begin{algorithm}[hptb]
	\caption{The calculation and storage of the reusable intermediate variables}
	\label{sgd_cufastertucker_ab}
	\footnotesize
	\vspace{.1cm}
	$\mathcal{G}\{parameter\}$: parameter in global memory.\\
	$\mathcal{C}\{parameter\}$: parameter in on-chip cache memory.\\
	$\mathcal{R}\{parameter\}$: parameter in register memory.\\
	$\textbf{Input}$: Initialized factor matrices $\textbf{A}^{(n)}$ $\in$ $\mathbb{R}^{I_{n}\times J_{n}}$, $n \in \{N\}$ and core matrices $\textbf{B}^{(n)}$ $\in$ $\mathbb{R}^{J_{n}\times R}$, $n \in \{N\}$.\\
	$\textbf{Output}$: Reusable intermediate variables $\textbf{a}^{(n)}_{i_{n}}\textbf{b}^{(n)}_{:,r}, n \in \{N\}, r \in \{R\}$.\\	
	\begin{algorithmic}[1]
		\FOR{$n$ from $1$ to $N$}
		\FOR{$r$ from $1$ to $R$}
		\STATE $\mathcal{C}\{\textbf{b}^{(n)}_{:,r}\}$ $\leftarrow$ $\mathcal{G}\{\textbf{b}^{(n)}_{:,r}\}$
		\ENDFOR	
		\FOR{$i_n$ from $1$ to $I_n$ (\emph{Worker Parallelization})}
		\STATE $\mathcal{R}\{\textbf{a}^{(n)}_{i_{n}}\}$ $\leftarrow$ $\mathcal{G}\{\textbf{a}^{(n)}_{i_{n}}\}$ 
		\FOR{$r$ from $1$ to $R$}
		\STATE $\mathcal{G}\{\textbf{a}^{(n)}_{i_{n}}\textbf{b}^{(n)}_{:,r}\}$ $\leftarrow$ $\mathcal{R}\{\textbf{a}^{(n)}_{i_{n}}\}$ $\cdot$ $\mathcal{C}\{\textbf{b}^{(n)}_{:,r}\}$
		\ENDFOR
		\ENDFOR
		\ENDFOR
	\end{algorithmic}
\end{algorithm}

\textbf{Memory Coalescing}:
The GPU's global memory is implemented with Dynamic Random Access Memory (DRAM), but DRAM is slow. 
Based on the parallelism of DRAM, we can achieve higher global memory access efficiency by optimizing the memory access pattern of threads, namely the Memory Coalescing technique. Memory Coalescing takes advantage of the fact that at any given point in time, threads in a warp are executing the same instructions, then optimal memory pattern will be achieved when all threads in a warp access global memory locations contiguously access mode. 
According to the algorithm characteristics of cuFasterTucker, 
all the variable matrices
$\textbf{A}^{(n)}$ $\in$ $\mathbb{R}^{I_{n}\times J_{n}}$, $n$ $\in$ $\{N\}$ and
$\textbf{B}^{(n)}$ $\in$ $\mathbb{R}^{J_{n}\times R}$, $n$ $\in$ $\{N\}$ 
are stored as the form of $\textbf{A}^{(n)}$ $\in$ $\mathbb{R}^{I_{n}\times J_{n}}$, $n$ $\in$ $\{N\}$
and $\textbf{B}^{(n)^{T}}$ $\in$ $\mathbb{R}^{R \times J_{n}}$, $n$ $\in$ $\{N\}$ 
to ensure that consecutive threads access consecutive memory addresses.

\begin{algorithm}[]
	\caption{Update factor matrices in cuFasterTucker}
	\label{sgd_cufastertucker_a}
	\vspace{.1cm}
	\footnotesize
	$\mathcal{G}\{parameter\}$: parameter in global memory.\\
	$\mathcal{C}\{parameter\}$: parameter in on-chip cache memory.\\
	$\mathcal{R}\{parameter\}$: parameter in register memory.\\
	$\textbf{Input}$: Sparse tensor $\mathcal{X}$ $\in$ $\mathbb{R}^{I_{1}\times\cdots\times I_{N}}$, 
	initialized factor matrices $\textbf{A}^{(n)}$ $\in$ $\mathbb{R}^{I_{n}\times J_{n}}$, $n \in \{N\}$ and core matrices $\textbf{B}^{(n)}$ $\in$ $\mathbb{R}^{J_{n}\times R}$, $n \in \{N\}$,
	reusable intermediate variables $\textbf{a}^{(n)}_{i_{n}}\textbf{b}^{(n)}_{:,r}, n \in \{N\}, r \in \{R\}$,
	learning rate $\gamma_{\textbf{A}}$, regularization parameter $\lambda_{\textbf{A}}$.\\
	$\textbf{Output}$: Factor matrices $\textbf{A}^{(n)}$, $n$ $\in$ $\{N\}$.\\	
	\begin{algorithmic}[1]
		\FOR{$n$ from $1$ to $N$}
		\FOR{$i_n$ from $1$ to $I^{'}_{n}$ (\emph{Worker Parallelization})}
		\FOR{$r$ from $1$ to $R$}
		\STATE $\mathcal{C}\{\textbf{b}^{(n)}_{:,r}\}$ $\leftarrow$ $\mathcal{G}\{\textbf{b}^{(n)}_{:,r}\}$
		\ENDFOR	
		\FOR{preorder traversal of subtensor $\mathcal{X}^{'}_{i_n}$}
		\STATE $d$ $=$ the depth of the subtensor $\mathcal{X}^{'}_{i_n}$.
		\STATE $n^{'}$ $=$ $(n+d)\%n$.
		\IF{$d$ $=$ $N-1$}
		\STATE $\mathcal{R}\{x_{i_1,\dots,i_{n^{'}},\dots,i_{N}}\}$ $\leftarrow$ $\mathcal{G}\{x_{i_1,\dots,i_{n^{'}},\dots,i_{N}}\}$ $-$ $\mathcal{G}\{\textbf{a}^{(n^{'})}_{i_{n^{'}}}\}$
		$\cdot$ $\mathcal{R}\{\textbf{B}^{(n)}\textbf{Q}^{(n)^{T}}\textbf{s}^{(n)^{T}}_{i_1,\dots,i_N}\}$
		\STATE $\mathcal{G}\{\textbf{a}^{(n^{'})}_{i_{n^{'}}}\}$ $\leftarrow$ $\mathcal{G}\{\textbf{a}^{(n^{'})}_{i_{n^{'}}}\}$ $-$ $\gamma_{\textbf{A}}$ $\cdot$
		$\big($ $\mathcal{R}\{x_{i_1,\dots,i_{n^{'}},\dots,i_{N}}\}$ $\cdot$ 
		$\mathcal{G}\{\textbf{a}^{(n^{'})}_{i_{n^{'}}}\}$ $\cdot$ $\mathcal{R}\{\textbf{B}^{(n)}\textbf{Q}^{(n)^{T}}\textbf{s}^{(n)^{T}}_{i_1,\dots,i_N}\}$ $+$
		$\lambda_{\textbf{A}}$ $\cdot$ $\mathcal{G}\{\textbf{a}^{(n^{'})}_{i_{n^{'}}}\}$ $\big)$
		\ELSIF{$d$ $=$ $N-2$}
		\STATE $\mathcal{R}\{\textbf{B}^{(n)}\textbf{Q}^{(n)^{T}}\textbf{s}^{(n)^{T}}_{i_1,\dots,i_N}\}$ $\leftarrow$ $0$
		\FOR{$r$ from $1$ to $R$}
		\STATE $\mathcal{R}\{\textbf{b}^{(n)}_{:,r}\textbf{s}^{(n)}_{i_1,\dots,i_N}\textbf{q}^{(n)}_{:,r}\}$ $\leftarrow$ $\mathcal{C}\{\textbf{b}^{(n)}_{:,r}\}$
		\FOR{$n^{''}$ from $1$ to $N$, $n^{''}$ $\neq$ $(n+N-1)\%n$}
		\STATE $\mathcal{R}\{\textbf{b}^{(n)}_{:,r}\textbf{s}^{(n)}_{i_1,\dots,i_N}\textbf{q}^{(n)}_{:,r}\}$ $\leftarrow$ $\mathcal{R}\{\textbf{b}^{(n)}_{:,r}\textbf{s}^{(n)}_{i_1,\dots,i_N}\textbf{q}^{(n)}_{:,r}\}$ $\cdot$ 
		$\mathcal{C}\{\textbf{a}^{(n^{''})}_{i_{n^{''}}}\textbf{b}^{(n^{''})}_{:,r}\}$
		\ENDFOR
		\STATE $\mathcal{R}\{\textbf{B}^{(n)}\textbf{Q}^{(n)^{T}}\textbf{s}^{(n)^{T}}_{i_1,\dots,i_N}\}$ $\leftarrow$
		$\mathcal{R}\{\textbf{B}^{(n)}\textbf{Q}^{(n)^{T}}\textbf{s}^{(n)^{T}}_{i_1,\dots,i_N}\}$ $+$
		$\mathcal{R}\{\textbf{b}^{(n)}_{:,r}\textbf{s}^{(n)}_{i_1,\dots,i_N}\textbf{q}^{(n)}_{:,r}\}$
		\ENDFOR
		\ELSE
		\FOR{$r$ from $1$ to $R$}
		\STATE $\mathcal{C}\{\textbf{a}^{(n^{'})}_{i_{n}^{'}}\textbf{b}^{(n^{'})}_{:,r}\}$ $\leftarrow$ $\mathcal{G}\{\textbf{a}^{(n^{'})}_{i_{n}}\textbf{b}^{(n^{'})}_{:,r}\}$ 
		\ENDFOR
		\ENDIF
		\ENDFOR
		\ENDFOR
		\ENDFOR
	\end{algorithmic}
\end{algorithm}
\textbf{Warp Shuffle}:
The warp shuffle instructions allows a thread to directly read the register values of other threads, 
as long as these threads are in the same warp. 
It is implemented through additional hardware support, 
which is better than shared memory for inter-thread communication, has lower latency, 
and does not consume additional memory resources to perform data exchange.
The warp shuffle instructions is commonly used to calculate dot product and sum in scientific computing, in cuFasterTucker we use to calculate $\textbf{a}^{(n)}_{i_{n}}\cdot\textbf{b}^{(n)}_{:,r}, i_n \in \{I_n\}, n \in \{N\}, r \in \{R\}$ and $\textbf{a}^{(n)}_{i_{n}}\cdot\textbf{B}^{(n)}\textbf{Q}^{(n)^{T}}\textbf{s}^{(n)^{T}}_{i_1,\dots,i_N}, i_n \in \{I_n\}, n \in \{N\}$.

\textbf{On-chip Cache}:
The current mainstream architecture of NVIDIA NGPU allows programmatic control over the caching behavior of the on-chip L1 cache. 
In cuFasterTucker, 
we use $\_\_ldg$ to put predictably reusable intermediate variables and commonly used variables into the on-chip L1 cache to improve memory access efficiency, 
such as $\textbf{a}^{(n)}_{i_{n}}\textbf{b}^{(n)}_{:,r}, i_n \in \{I_n\}, n \in \{N\}, r \in \{R\}$ and $\textbf{b}^{(n)}_{:,r}, n \in \{N\}, r \in \{R\}$ stored in global memory.
In practice, the GPU's cache scheduling also works well.

\textbf{Shared Memory}:
In the current mainstream architecture of nvidia gpus, 
shared memory is put into the on-chip L1 cache, 
which is much faster than global memory. 
At the same time registers on the gpu are not suitable for storing contiguous vectors. 
Therefore cuFasterTucker uses shared memory to store reusable intermediate vectors and use them for the next process.
In the update core matrices module, 
we use shared memory to store $\textbf{s}^{(n)}_{i_1,\dots,i_N}\textbf{q}^{(n)}_{:,r}, n \in \{N\}, r \in \{R\}$.

\textbf{Register}:
The storage structure of each variable of cuFasterTucker is clearly allocated, 
and all variables occupy only a small amount of storage space, 
that is, the registers in the gpu are completely sufficient for cuFasterTucker.
Due to the structure of B-CSF, 
the index and value of tensor $\mathcal{X}$ are read and put into registers for the next process without re-reading from global memory.
In addition, a register of each thread can store a scalar, 
and a thread group can store a vector. 
The premise of such storage is that each thread is only responsible for reading, writing and computing its own registers.
In cuFasterTucker, vectors $\textbf{a}^{(n)}_{i_{n}}, i_{n} \in \{I_{n}\}, n \in \{N\}$ and $\textbf{B}^{(n)}\textbf{Q}^{(n)^{T}}\textbf{s}^{(n)^{T}}_{i_1,\dots,i_N}, n \in \{N\}$ are all stored in registers in this form.

\subsection{Overview} \label{Section Overview}
cuFasterTucker is mainly composed of three parts, 
the first part is the calculation and storage of the reusable intermediate variables $\textbf{a}^{(n)}_{i_{n}}\textbf{b}^{(n)}_{:,r}, n \in \{N\}, r \in \{R\}$, 
the second part is the update factor matrices, 
and the third part is the update core matrices. 
Updating the factor matrices and updating the core matrices can be performed independently or simultaneously.
But before that, the reusable intermediate variable $\textbf{a}^{(n)}_{i_{n}}\textbf{b}^{(n)}_{:,r}, n \in \{N\}, r \in \{R\}$ must be computed and stored.
We use three algorithms to separately describe in detail how these three parts are implemented in parallel on the GPU.
The worker parallelization in the algorithms has been marked, and each process is thread parallelization.

Algorithm \ref{sgd_cufastertucker_ab} describes the process of computing and storing $\textbf{a}^{(n)}_{i_{n}}\cdot\textbf{b}^{(n)}_{:,r}, i_n \in \{I_n\}, n \in \{N\}, r \in \{R\}$.
For each order $n$, 
the repeatedly used $\textbf{b}^{(n)}_{:,r}, r \in \{R\}$ is placed in the on-chip cache at first to improve the memory access efficiency during the calculation process (Algorithm \ref{sgd_cufastertucker_ab}, Line 3).
Next, multiple workers process $\textbf{a}^{(n)}_{i_{n}}, i_{n} \in \{I_{n}\}$ in parallel and put $\textbf{a}^{(n)}_{i_{n}}$ into registers to improve the memory access efficiency of $\textbf{a}^{(n)}_{i_{n}}$ for the next calculation of $\textbf{a}^{(n)}_{i_{n}}\cdot\textbf{b}^{(n)}_{:,r}, r \in \{R\}$ (Algorithm \ref{sgd_cufastertucker_ab}, Line 6).
Finally, the calculated $\textbf{a}^{(n)}_{i_{n}}\cdot\textbf{b}^{(n)}_{:,r}, i_n \in \{I_n\}, r \in \{R\}$ is stored in global memory for use when updating the factor matrices or core matrices (Algorithm \ref{sgd_cufastertucker_ab}, Line 8).

\begin{algorithm}[]
	\caption{Update core matrices in cuFasterTucker}
	\label{sgd_fastertucker_b}
	\footnotesize
	\vspace{.1cm}
	$\mathcal{G}\{parameter\}$: parameter in global memory.\\
	$\mathcal{C}\{parameter\}$: parameter in on-chip cache memory.\\
	$\mathcal{S}\{parameter\}$: parameter in shared memory.\\
	$\mathcal{R}\{parameter\}$: parameter in register memory.\\
	$\textbf{Input}$: Sparse tensor $\mathcal{X}$ $\in$ $\mathbb{R}^{I_{1}\times\cdots\times I_{N}}$, 
	initialized factor matrices $\textbf{A}^{(n)}$ $\in$ $\mathbb{R}^{I_{n}\times J_{n}}$, $n \in \{N\}$ and core matrices $\textbf{B}^{(n)}$ $\in$ $\mathbb{R}^{J_{n}\times R}$, $n \in \{N\}$,
	learning rate $\gamma_{\textbf{B}}$,
	regularization parameter $\lambda_{\textbf{B}}$.\\
	$\textbf{Output}$: Core matrices $\textbf{B}^{(n)}$, $n$ $\in$ $\{N\}$.\\	
	\begin{algorithmic}[1]
		\FOR{$n$ from $1$ to $N$}
		\FOR{$r$ from $1$ to $R$}
		\STATE $\mathcal{G}\{$ $The$ $gradient$ $of$ $\textbf{b}^{(n)}_{:,r}\}$ $\leftarrow$ $0$
		\ENDFOR	
		\FOR{$i_n$ from $1$ to $I^{'}_{n}$ (\emph{Worker Parallelization})}
		\FOR{$r$ from $1$ to $R$}
		\STATE $\mathcal{C}\{\textbf{b}^{(n)}_{:,r}\}$ $\leftarrow$ $\mathcal{G}\{\textbf{b}^{(n)}_{:,r}\}$
		\ENDFOR	
		\FOR{preorder traversal of subtensor $\mathcal{X}^{'}_{i_n}$}
		\STATE $d$ $=$ the depth of the subtensor $\mathcal{X}^{'}_{i_n}$.
		\STATE $n^{'}$ $=$ $(n+d)\%n$.
		\IF{$d$ $=$ $N-1$}	
		\STATE $\mathcal{R}\{x_{i_1,\dots,i_{n^{'}},\dots,i_{N}}\}$ $\leftarrow$ $\mathcal{G}\{x_{i_1,\dots,i_{n^{'}},\dots,i_{N}}\}$ $-$ $\mathcal{G}\{\textbf{a}^{(n^{'})}_{i_{n^{'}}}\}$
		$\cdot$ $\mathcal{R}\{\textbf{B}^{(n)}\textbf{Q}^{(n)^{T}}\textbf{s}^{(n)^{T}}_{i_1,\dots,i_N}\}$
		\FOR{$r$ from $1$ to $R$}
		\STATE $\mathcal{G}\{$ $The$ $gradient$ $of$ $\textbf{b}^{(n)}_{:,r}\}$ $\leftarrow$ 
		$\mathcal{G}\{$ $The$ $gradient$ $of$ $\textbf{b}^{(n)}_{:,r}\}$ $+$ 
		$\mathcal{R}\{x_{i_1,\dots,i_{n^{'}},\dots,i_{N}}\}$ $\cdot$ 
		$\mathcal{G}\{\textbf{a}^{(n^{'})}_{i_{n^{'}}}\}$ $\cdot$ 
		$\mathcal{S}\{\textbf{s}^{(n)}_{i_1,\dots,i_N}\textbf{q}^{(n)}_{:,r}\}$
		\ENDFOR	
		\ELSIF{$d$ $=$ $N-2$}
		\STATE $\mathcal{R}\{\textbf{B}^{(n)}\textbf{Q}^{(n)^{T}}\textbf{s}^{(n)^{T}}_{i_1,\dots,i_N}\}$ $\leftarrow$ $0$
		\FOR{$r$ from $1$ to $R$}
		\STATE $\mathcal{S}\{\textbf{s}^{(n)}_{i_1,\dots,i_N}\textbf{q}^{(n)}_{:,r}\}$ $\leftarrow$ $1.0$
		\FOR{$n^{''}$ from $1$ to $N$, $n^{''}$ $\neq$ $(n+N-1)\%n$}
		\STATE $\mathcal{S}\{\textbf{s}^{(n)}_{i_1,\dots,i_N}\textbf{q}^{(n)}_{:,r}\}$ $\leftarrow$ $\mathcal{S}\{\textbf{s}^{(n)}_{i_1,\dots,i_N}\textbf{q}^{(n)}_{:,r}\}$ $\cdot$ 
		$\mathcal{C}\{\textbf{a}^{(n^{''})}_{i_{n^{''}}}\textbf{b}^{(n^{''})}_{:,r}\}$
		\ENDFOR
		\STATE $\mathcal{R}\{\textbf{B}^{(n)}\textbf{Q}^{(n)^{T}}\textbf{s}^{(n)^{T}}_{i_1,\dots,i_N}\}$ $\leftarrow$
		$\mathcal{R}\{\textbf{B}^{(n)}\textbf{Q}^{(n)^{T}}\textbf{s}^{(n)^{T}}_{i_1,\dots,i_N}\}$ $+$
		$\mathcal{S}\{\textbf{s}^{(n)}_{i_1,\dots,i_N}\textbf{q}^{(n)}_{:,r}\}$ $\cdot$ $\mathcal{C}\{\textbf{b}^{(n)}_{:,r}\}$
		\ENDFOR
		\ELSE
		\FOR{$r$ from $1$ to $R$}
		\STATE $\mathcal{C}\{\textbf{a}^{(n^{'})}_{i_{n}}\textbf{b}^{(n^{'})}_{:,r}\}$ $\leftarrow$ $\mathcal{G}\{\textbf{a}^{(n^{'})}_{i_{n}}\textbf{b}^{(n^{'})}_{:,r}\}$ 
		\ENDFOR
		\ENDIF
		\ENDFOR
		\ENDFOR	
		\FOR{$r$ from $1$ to $R$}
		\STATE $\mathcal{G}\{\textbf{b}^{(n)}_{:,r}\}$ $\leftarrow$ $\mathcal{G}\{\textbf{b}^{(n)}_{:,r}\}$ 
		$-$ $\gamma_{\textbf{B}}$ $\cdot$
		$\big($ $\mathcal{G}\{$ $The$ $gradient$ $of$ $\textbf{b}^{(n)}_{:,r}\}$ $/$ $|\Omega|$ $+$
		$\lambda_{\textbf{B}}$ $\cdot$ $\mathcal{G}\{\textbf{b}^{(n)}_{:,r}\}$ $\big)$
		\ENDFOR	
		\ENDFOR	
	\end{algorithmic}
\end{algorithm}

Algorithm \ref{sgd_cufastertucker_a} describes the process of updating $\textbf{a}^{(n)}_{i_{n}}, i_{n} \in \{I_{n}\}, n \in \{N\}$.
For each order $n$, 
multiple workers process $I^{'}_{n}$ sub-tensors $\mathcal{X}^{'}_{i_n}$ in parallel (Algorithm \ref{sgd_cufastertucker_a}, line 2).
The repeatedly used $\textbf{b}^{(n)}_{:,r}, r \in \{R\}$ is placed in the on-chip cache at first to improve the memory access efficiency during the calculation process (Algorithm \ref{sgd_cufastertucker_a}, line 4).
Next, each subtensor $\mathcal{X}^{'}_{i_n}$ is preorder traversed by a worker (Algorithm \ref{sgd_cufastertucker_a}, line 6).
If the node is not a leaf node or the parent node of the leaf node (Algorithm \ref{sgd_cufastertucker_a}, line 21), 
the corresponding intermediate variable $\textbf{a}^{(n^{'})}_{i_{n}^{'}}\textbf{b}^{(n^{'})}_{:,r}$ is put into the on-chip cache (Algorithm \ref{sgd_cufastertucker_a}, line 23), 
which provides memory access acceleration for the subsequent computation.
If the node is the parent node of the leaf node (Algorithm \ref{sgd_cufastertucker_a}, line 12), 
the shared intermediate variable $\textbf{B}^{(n)}\textbf{Q}^{(n)^{T}}\textbf{s}^{(n)^{T}}_{i_1,\dots,i_N}$ is calculated and put into the register (Algorithm \ref{sgd_cufastertucker_a}, lines 13-20).
If the node is a leaf node (Algorithm \ref{sgd_cufastertucker_a}, line 9), 
the corresponding $\textbf{a}^{(n^{'})}_{i_{n^{'}}}$ is updated (Algorithm \ref{sgd_cufastertucker_a}, lines 10-11).
Each time a order is processed, $\textbf{a}^{(n)}_{i_{n}}\textbf{b}^{(n)}_{:,r}, n \in \{N\}, r \in \{R\}$ in global memory needs to be updated (Algorithm \ref{sgd_cufastertucker_ab}, Lines 2-10). 
Note that since the B-CSF tensor storage format is used, when dealing with the $n$th order, $\textbf{a}^{(n)}_{i_{n}}\textbf{b}^{(n)}_{:,r}, n=(n+N-1)\%n, i_n \in \{I_n\}, r \in \{R\}$ is updated.

Algorithm \ref{sgd_fastertucker_b} describes the process of updating $\textbf{b}^{(n)}_{:,r}, n \in \{N\}, r \in \{R\}$.
Since the core matrices is required by all workers, and the scale of the core matrices in practical applications is usually small, 
it is not suitable to update the core matrices in parallel in real time.
We accumulate the gradients of each set of non-zero elements and update the core matrices uniformly at the end.
For each order $n$, 
store the gradient of $\textbf{b}^{(n)}_{:,r}\}, r \in \{R\}$ in global memory at first (Algorithm \ref{sgd_fastertucker_b}, line 3).
Next, multiple workers process $I^{'}_{n}$ sub-tensors $\mathcal{X}^{'}_{i_n}$ in parallel (Algorithm \ref{sgd_fastertucker_b}, line 5).
Then, the repeatedly used $\textbf{b}^{(n)}_{:,r}, r \in \{R\}$ is placed in the on-chip cache to improve the memory access efficiency during the calculation process (Algorithm \ref{sgd_fastertucker_b}, line 7).
Each subtensor $\mathcal{X}^{'}_{i_n}$ is preorder traversed by a worker (Algorithm \ref{sgd_fastertucker_b}, line 9).
If the node is not a leaf node or the parent node of the leaf node (Algorithm \ref{sgd_fastertucker_b}, line 26), 
the corresponding intermediate variable $\textbf{a}^{(n^{'})}_{i_{n}^{'}}\textbf{b}^{(n^{'})}_{:,r}$ is put into the on-chip cache (Algorithm \ref{sgd_fastertucker_b}, lines 27-29), 
which provides memory access acceleration for the subsequent computation.
If the node is the parent node of the leaf node (Algorithm \ref{sgd_fastertucker_b}, line 17), 
the shared intermediate variable $\textbf{s}^{(n)}_{i_1,\dots,i_N}\textbf{q}^{(n)}_{:,r}, r \in \{R\}$ is calculated and put into the shared memory (Algorithm \ref{sgd_fastertucker_b}, lines 18-25).
If the node is a leaf node (Algorithm \ref{sgd_fastertucker_b}, line 12), 
the gradient of $\textbf{b}^{(n)}_{:,r}\}, r \in \{R\}$ is computed and accumulated into global memory.
After all gradients are accumulated, the $\textbf{b}^{(n)}_{:,r}\}, r \in \{R\}$ 
updated (Algorithm \ref{sgd_fastertucker_b}, lines 13-16).
Same as when updating factor matrices, each time a order is processed, $\textbf{a}^{(n)}_{i_{n}}\textbf{b}^{(n)}_{:,r}, n \in \{N\}, r \in \{R\}$ in global memory needs to be updated (Algorithm \ref{sgd_cufastertucker_ab}, Lines 2-10). 
And when dealing with the $n$th order, $\textbf{a}^{(n)}_{i_{n}}\textbf{b}^{(n)}_{:,r}, n=(n+N-1)\%n, i_n \in \{I_n\}, r \in \{R\}$ is updated.

\section{Experiments} \label{Section Experiments}

We present experimental results to answer the following questions.

\begin{enumerate}
	
	\item \emph{Time for a single iteration}. 
	How well does cuFasterTucker and its contrasting algorithms perform in terms of iteration speed? 
	Is it the same efficiency for tensor storage format using COO or CSF?
	Does the extraction of reusable intermediate variables and shared invariant intermediate Variables improve the efficiency of the algorithm?
	
	\item \emph{Real-world accuracy}. How accurate do cuFasterTucker and its contrasting algorithms on real-world datasets?
	
	\item \emph{Adaptability of high-order tensors}. How well does cuFasterTucker and its contrasting algorithms perform on high-order datasets?
	
	\item \emph{Adaptability of tensor sparsity}. Does the sparsity of tensors have an effect on cufast and its contrasting algorithms?
	
\end{enumerate}

We describe the datasets and experimental settings in Section \ref{Section Experimental Settings}, 
and answer the questions in Sections \ref{Section Time for a single iteration} to \ref{Section Adaptability of tensor sparsity}.

\begin{table}[htbp]
	\centering
	\footnotesize
	\setlength{\abovecaptionskip}{0pt}
	\caption{Real World Datasets}
	\begin{tabular}{c|cc}
		\hline
		\hline
						& Netflix        & Yahoo!Music       \\
		\hline
		$I_1$           & 480, 189       & 1, 000, 990       \\
		$I_2$           & 17, 770        & 624, 961          \\
		$I_3$           & 2, 182         & 3, 075            \\
		$|\Omega|$      & 99, 072, 112   & 250, 272, 286     \\
		$|\Gamma|$      & 1, 408, 395    & 2, 527, 989       \\
		Max Value       & 5              & 5                 \\
		Min Value       & 1              & 0.025             \\
		\hline
		\hline
	\end{tabular}
	\label{data_sets_real_word}
\end{table}

\begin{table}[htbp]
	\centering
	\footnotesize
	\setlength{\abovecaptionskip}{0pt}
	\caption{Synthetic Datasets}
	\begin{tabular}{c|cc}
		\hline
		\hline
						& Synthetic(Order)   & Synthetic(Sparsity)       \\
		\hline
		$order$         & 3,4,5,6,7,8,9,10   & 3                         \\
		$I$             & 10, 000            & 1, 000                    \\
		$|\Omega|$      & 100M               & 20M,40M,60M,80M,100M      \\
		Max Value       & 5                  & 5                         \\
		Min Value       & 1                  & 1                         \\
		\hline
		\hline
	\end{tabular}
	\label{data_sets_synthetic}
\end{table}

\subsection{Experimental Settings} \label{Section Experimental Settings}

\begin{enumerate}
	
	\item \emph{Datasets}: We use both real-world and synthetic datasets to evaluate cuFasterTucker and its contrasting algorithms. 
	For real-world datasets, we use 
	Netflix\footnotemark[1] \footnotetext[1]{https://www.netflixprize.com/} 
	and Yahoo!music \footnotemark[2] \footnotetext[2]{https://webscope.sandbox.yahoo.com/}.
	Netflix is movie rating data which consist of (user, movie, time, rating).
	Yahoo!music is music rating data which consist of (user, music, time, rating).
	Using Netflix as the standard, we normalize all values of the Yahoo!music dataset between $0$ to $5$. 
	For synthetic datasets, we create two kinds of random tensors.
	One contains $8$ tensors with orders from $3$ to $10$, 
	while other parameters remain the same. 
	These tensors are used to examine the performance of cuFasterTucker and its contrasting algorithms on high-order tensors.
	The other contains $5$ tensors with the number of non-zero elements ranging from $20$ million to $100$ million, 
	and other parameters remain the same. 
	Since we set the order to $3$ and the length of each order is $1000$, 
	the sparsity of these tensors are $2\%$, $4\%$, $6\%$, $8\%$ and $10\%$ respectively.
	These tensors are used to examine the performance of cuFasterTucker and its contrasting algorithms on tensors of different sparsity.
	Tables \ref{data_sets_real_word} and \ref{data_sets_synthetic} describe the real-world datasets and synthetic datasets used in the experiments, respectively.
	
	\item \emph{Contrasting algorithms}: We compare cuFasterTucker and its variants with the state-of-the-art FasterTucker factorization algorithm cuFastTucker and other parallel sparse Tucker decomposition algorithms. As far as we know, there is only one FastTucker factorization algorithm.
	Descriptions of all methods are given as follows:
	
	\begin{itemize}
		
		\item \textbf{P-Tucker\cite{oh2018scalable}:} A scalable Tucker decomposition method for sparse tensors.
		
		\item \textbf{Vest\cite{park2021vest}:} A tensor decomposition method for large partially observable data to output a very sparse core tensor and factor matrices.
		
		\item \textbf{SGD\_\_Tucker\cite{li2020sgd}:} A Novel Stochastic Optimization Strategy for Parallel Sparse Tucker Decomposition.
				
		\item \textbf{ParTi\cite{parti}:} A fast essential sparse tensor operations and tensor decompositions method on multicore CPU and GPU architectures.
		
		\item \textbf{GTA\cite{oh2019high}:} A general framework for Tucker decomposition on heterogeneous platforms.
		
		\item \textbf{cuTucker\cite{li2022cu_fasttucker}:} A parallel sparse Tucker decomposition algorithm on the CUDA platform.
						
		\item \textbf{cuFastTucker\cite{li2022cu_fasttucker}:} A parallel sparse FastTucker decomposition algorithm on the CUDA platform with polynomial computational complexity.
		
		\item \textbf{cuFasterTucker\_COO:} A cuFastertucker algorithm, 
		which only reduces the computation of reusable intermediate variables, 
		uses the same tensor storage format COO as cuFastTucker.
		
		\item \textbf{cuFasterTucker\_B-CSF:} A cuFasterTucker algorithm, 
		which only reduces the computation of reusable intermediate variables, 
		uses B-CSF, a tensor storage format that is not identical to cuFastTucker.
		
		\item \textbf{cuFasterTucker:} The complete cuFasterTucker algorithm, which reduces the computation of reusable intermediate variables and shared intermediate variables.
		
	\end{itemize}

	\item \emph{Environment:} cuFasterTucker is implemented in C/C++ with CUDA.
	For its contrasting algorithm cuFastTucker,
	we use the original implementations provided by the authors.		
	The experiments of P-Tucker, Vest and SGD\_\_Tucker are ran on a \textbf{Intel Core i7-12700K CPU} with 64GB RAM.
	The experiments of ParTi, GTA, cuTucker, cuFastTucker, cuFasterTucker\_COO, cuFasterTucker\_B-CSF and cuFasterTucker are ran on a \textbf{NVIDIA GeForce RTX 3080Ti GPU} with 12GB graphic memory.
	We set the fiber threshold to $128$ for the B-CSF tensor storage format, which is considered to have the best performance.
	We set the maximum number of iterations to $50$ and report the average time for a single iteration.

\end{enumerate}

\subsection{Time for a single iteration} \label{Section Time for a single iteration}
We use real-world datasets to evaluate the running speed of cuFasterTucker and its contrasting algorithms, 
i.e. the time required for a single iteration.
In order to facilitate contrasting and maximize algorithm efficiency, all algorithms take $J_n=32, n \in \{N\}$ and $R=32$.
However, most of the advanced sparse Tucker decomposition algorithms can not adapt to the HOHDST.
Table \ref{The single iteration time or error of sparse Tucker decomposition algorithms} shows the single iteration time of these algorithms or the reason why they cannot work.
We relax the conditions appropriately. 
When $J_n=16, n \in \{N\}$, it takes $4829.497819$ seconds and $10816.486406$ seconds for Vest to update the factor matrices on Netflix and Yahoo!Music datasets, respectively.
And it takes $9492.699654$ seconds and $29746.486010$ seconds for Vest to update the core matrices on Netflix and Yahoo!Music datasets, respectively.
When $J_n=8, n \in \{N\}$, ParTi can run on the Yahoo!Music dataset, and its single iteration time is $54.861532$ seconds. 
In addition, the preprocessing time of ParTi on Netflix and Yahoo!Music datasets is $249.978767$ seconds and $646.884534$ seconds respectively.
When $J_n=16, n \in \{N\}$, GTA can run on the Netflix dataset, and the single iteration time is $243.797254$ seconds.
And when $J_n=8, n \in \{N\}$, GTA can run on the Yahoo!Music dataset, and the single iteration time is $22.904287$ seconds.
It can be seen that the above algorithms cannot meet our requirements for processing HOHDST.

\begin{table}[htbp]
	\centering
	\footnotesize
	\setlength{\abovecaptionskip}{0pt}
	\caption{The single iteration time (seconds) or error of sparse Tucker decomposition algorithms}
	\begin{tabular}{c|ccc}
		\hline \hline
		Algorithm                     & Netflix           & Yahoo!Music      \\
		\hline
		P-Tucker(Factor)              & 2886.727982       & 7381.461875      \\
		Vest(Factor)                  & out of time       & out of time      \\
		SGD\_\_Tucker(Factor)         & 609.306086        & 1354.774812      \\
		ParTi(Factor)                 & 67.537985         & out of memory    \\
		GTA(Factor)                   & out of memory     & out of memory    \\
		cuTucker(Factor)              & 64.648387         & 163.867682       \\
		\hline
		Vest(Core)                    & out of time       & out of time      \\
		cuTucker(Core)                & 63.404410         & 161.934037       \\
		\hline \hline
	\end{tabular}
	\label{The single iteration time or error of sparse Tucker decomposition algorithms}
\end{table}

In Table \ref{Speed comparison based on cuFastTucker}, the average time(seconds) for a single iteration of each sparse Fast Tucker decomposition algorithm are presented.
Due to the storage and invocation of reusable intermediate variables, cuFasterTucker\_COO achieves a speedup of more than $3.0X$ in updating factor matrices and updating core matrices compared to cuFastTucker.
Compared with cuFasterTucker\_COO, cuFasterTucker\_B-CSF is only a tensor storage format change, but therefore cuFasterTucker\_B-CSF has obtained a speedup of $2.0X$ to $5.0X$ relative to cuFastTucker in updating the factor matrices and the core matrices, respectively.
However, compared with cuFasterTucker\_B-CSF, there is a significant difference in the increase of the speedup ratio relative to cuFastTucker obtained by cuFasterTucker when updating the factor matrices and updating the core matrices.
When updating the factor matrices, cuFasterTucker has a speedup growth of $7.0X$ to $8.0X$, while updating the core matrices is only about $1.0X$.
This is because the proportion of shared intermediate variables when updating the factor matrices is much larger than updating the core matrices.
In fact, cuFastTucker and cuFasterTucker can run $10$ times the size of the Netflix dataset in 12GB of graphic memory.

\begin{table}[htbp]
	\centering
	\footnotesize
	\setlength{\abovecaptionskip}{0pt}
	\caption{Speedup comparison on the baseline cuFastTucker}
	\begin{tabular}{c|ccc}
		\hline \hline
		Algorithm                     & Netflix           & Yahoo!Music      \\
		\hline
		cuFastTucker(Factor)          & 4.558734          & 11.557935        \\
		cuFasterTucker\_COO(Factor)   & 1.385437(3.29X)   & 3.536883(3.27X)  \\
		cuFasterTucker\_B-CSF(Factor) & 0.534161(8.53X)   & 1.322512(8.74X)  \\
		cuFasterTucker(Factor)        & 0.294704(15.47X)  & 0.787373(14.68X) \\
		\hline
		cuFastTucker(Core)            & 6.044708          & 15.405755        \\
		cuFasterTucker\_COO(Core)     & 1.947172(3.10X)   & 4.976965(3.10X)  \\
		cuFasterTucker\_B-CSF(Core)   & 0.998262(6.06X)   & 2.505620(6.15X)  \\
		cuFasterTucker(Core)          & 0.835289(7.24X)   & 2.187568(7.04X)  \\
		\hline \hline
	\end{tabular}
	\label{Speed comparison based on cuFastTucker}
\end{table}

\begin{figure*}[htbp]
	\centering
	\subfigure[RMSE on Netflix]{
		\label{convergence(a)}
		\includegraphics[width=3in]{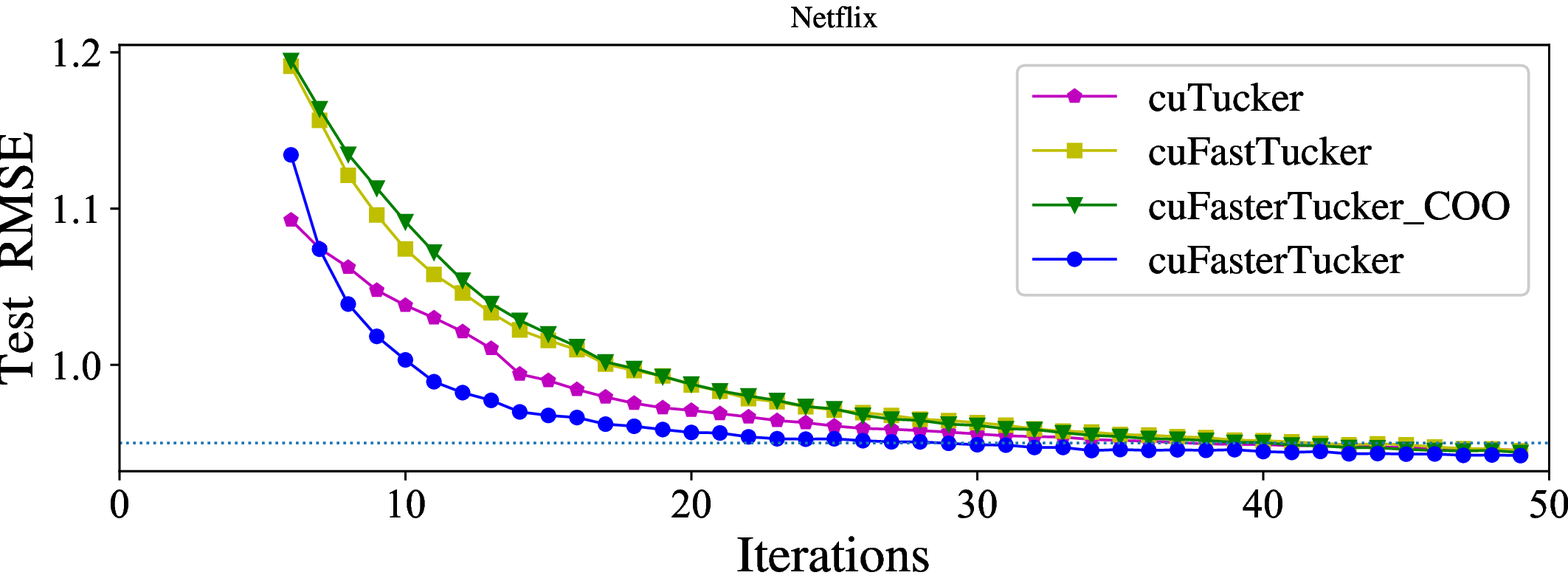}}
	~
	\subfigure[MAE on Netflix]{
		\label{convergence(b)}
		\includegraphics[width=3in]{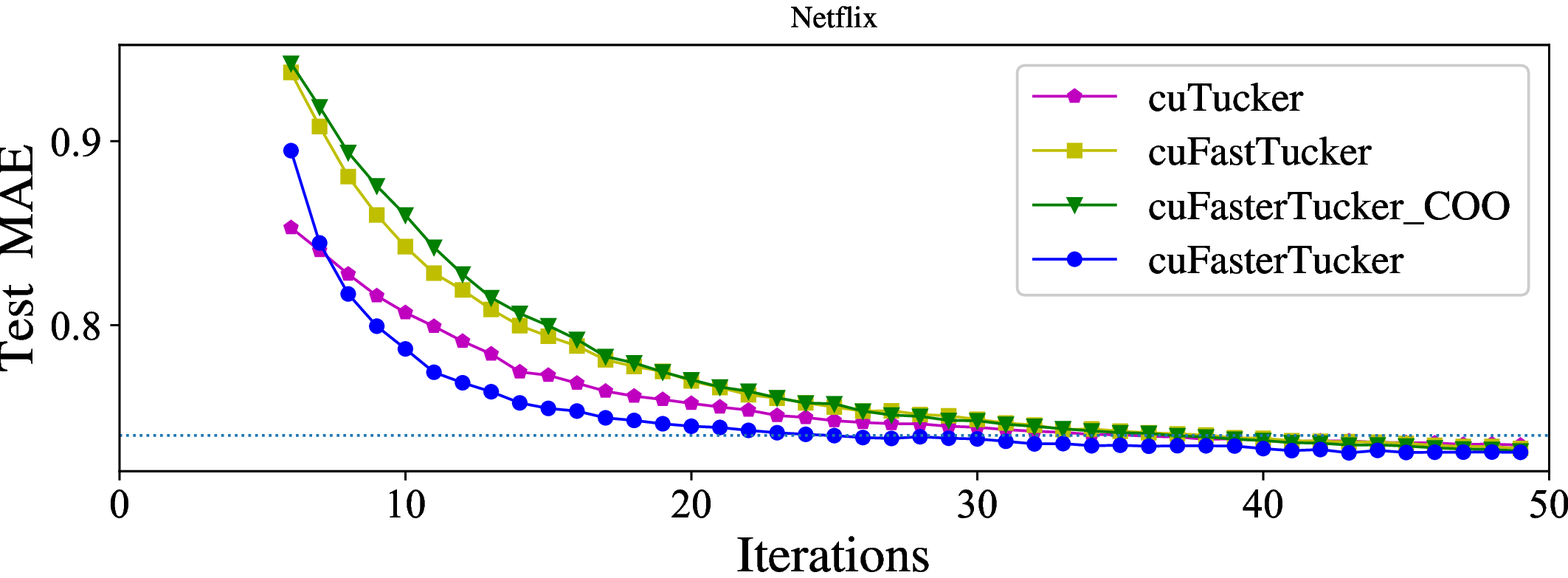}}
	~
	\subfigure[RMSE on Yahoo!Music]{
		\label{convergence(c)}
		\includegraphics[width=3in]{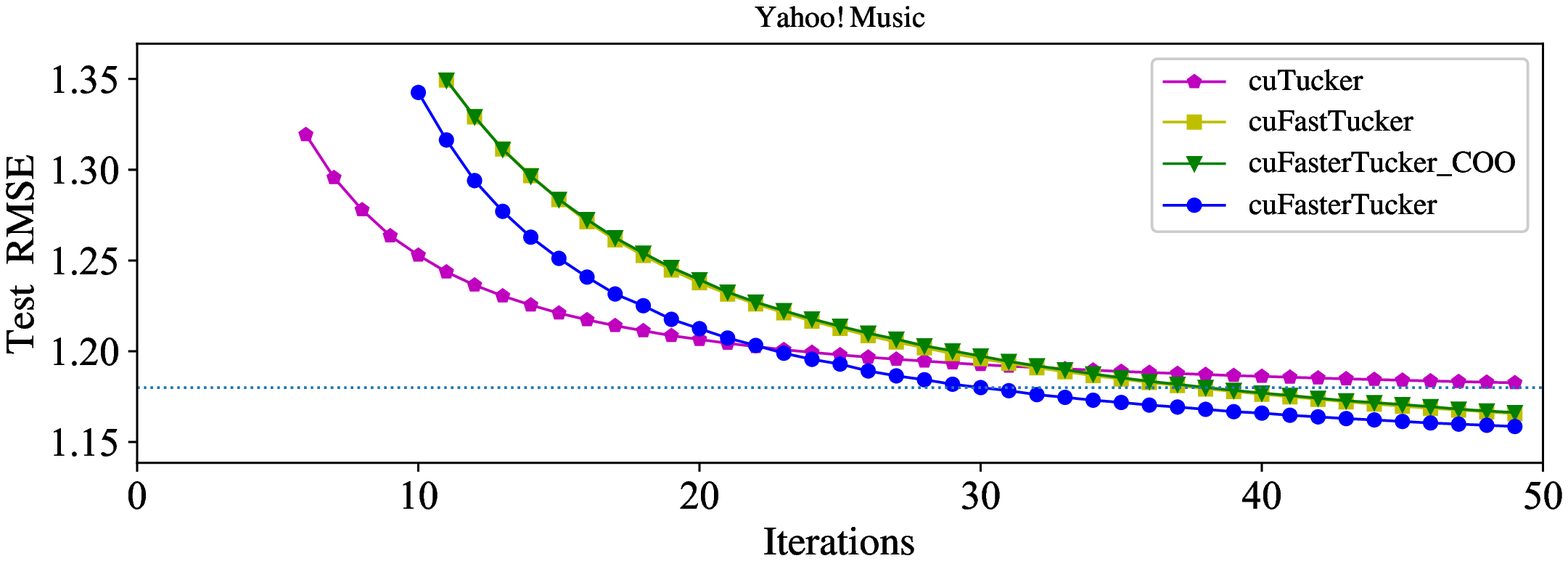}}
	~
	\subfigure[MAE on Yahoo!Music]{
		\label{convergence(d)}
		\includegraphics[width=3in]{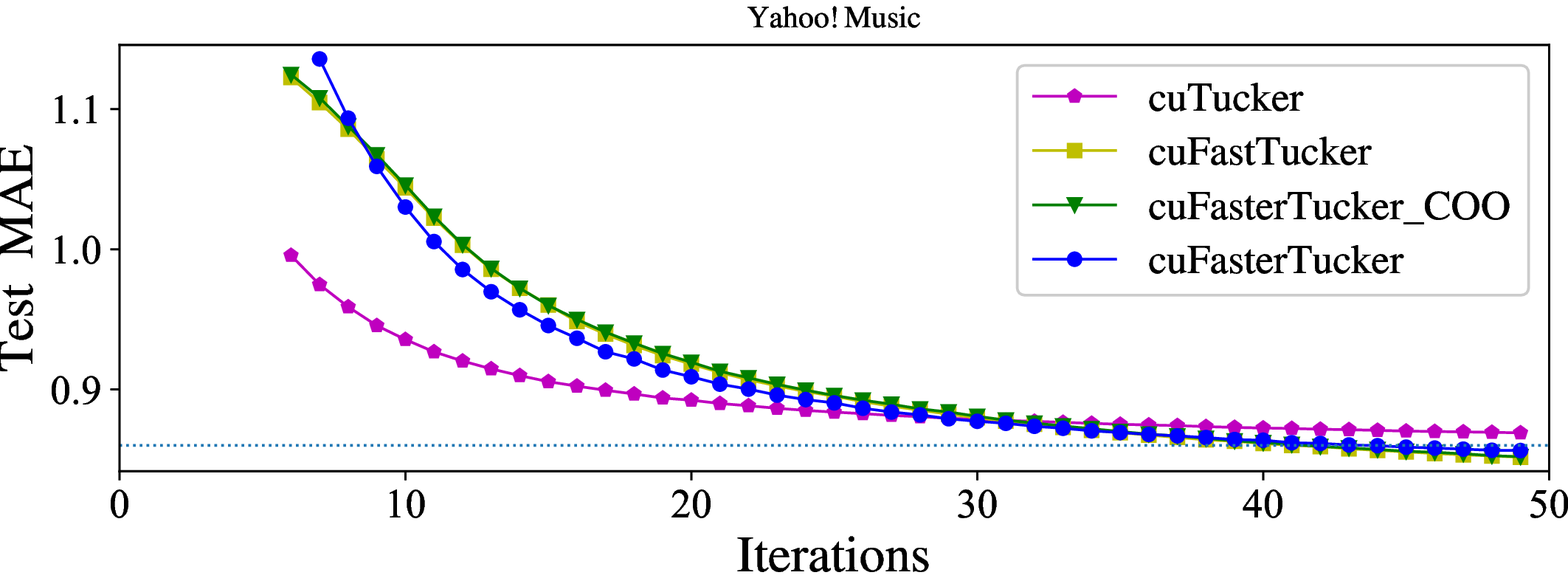}}
	\caption{Convergence curves of cuFasterTucker and its contrasting algorithm, they all set $J_n=32, n \in \{N\}$ and $R=32$.}
	\label{convergence}
\end{figure*}

\begin{figure*}[htbp]
	\centering
	\subfigure[Scalability on Synthesis Datasets]{
		\label{synthetic(a)}
		\includegraphics[width=2in]{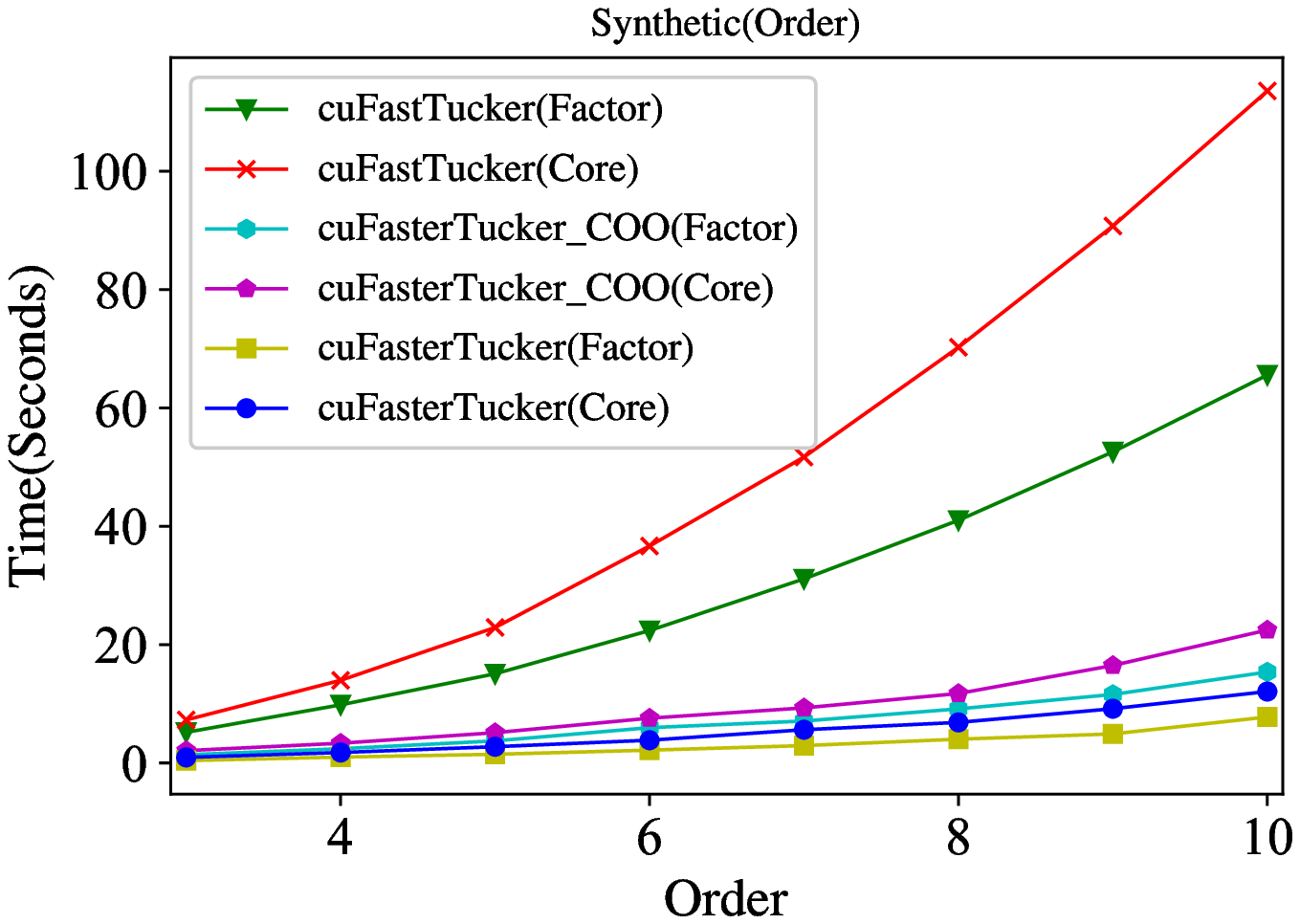}}
	~
	\subfigure[Speedup on Netflix]{
		\label{synthetic(b)}
		\includegraphics[width=2in]{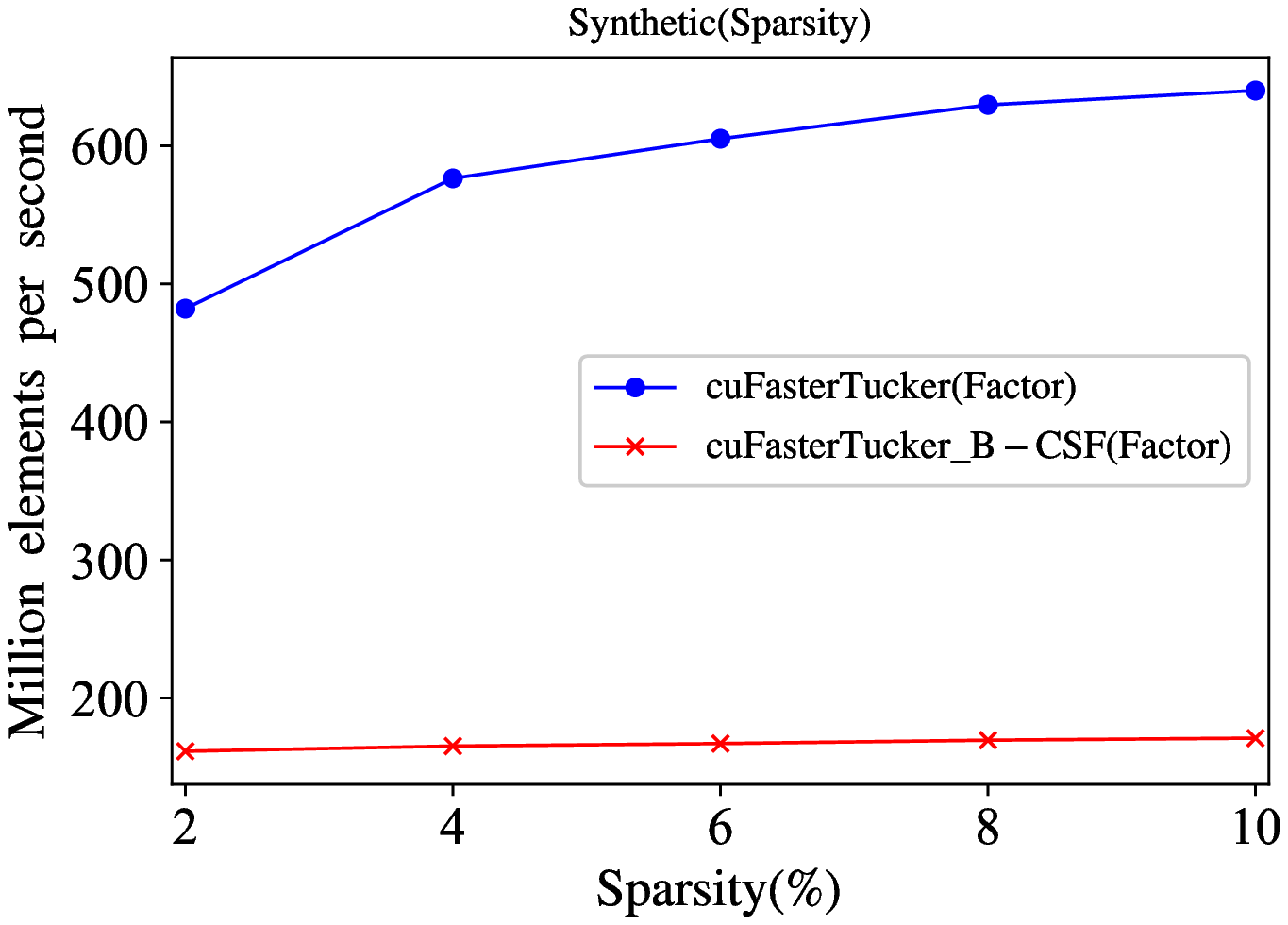}}
	~
	\subfigure[Speedup on Yahoo!Music]{
		\label{synthetic(c)}
		\includegraphics[width=2in]{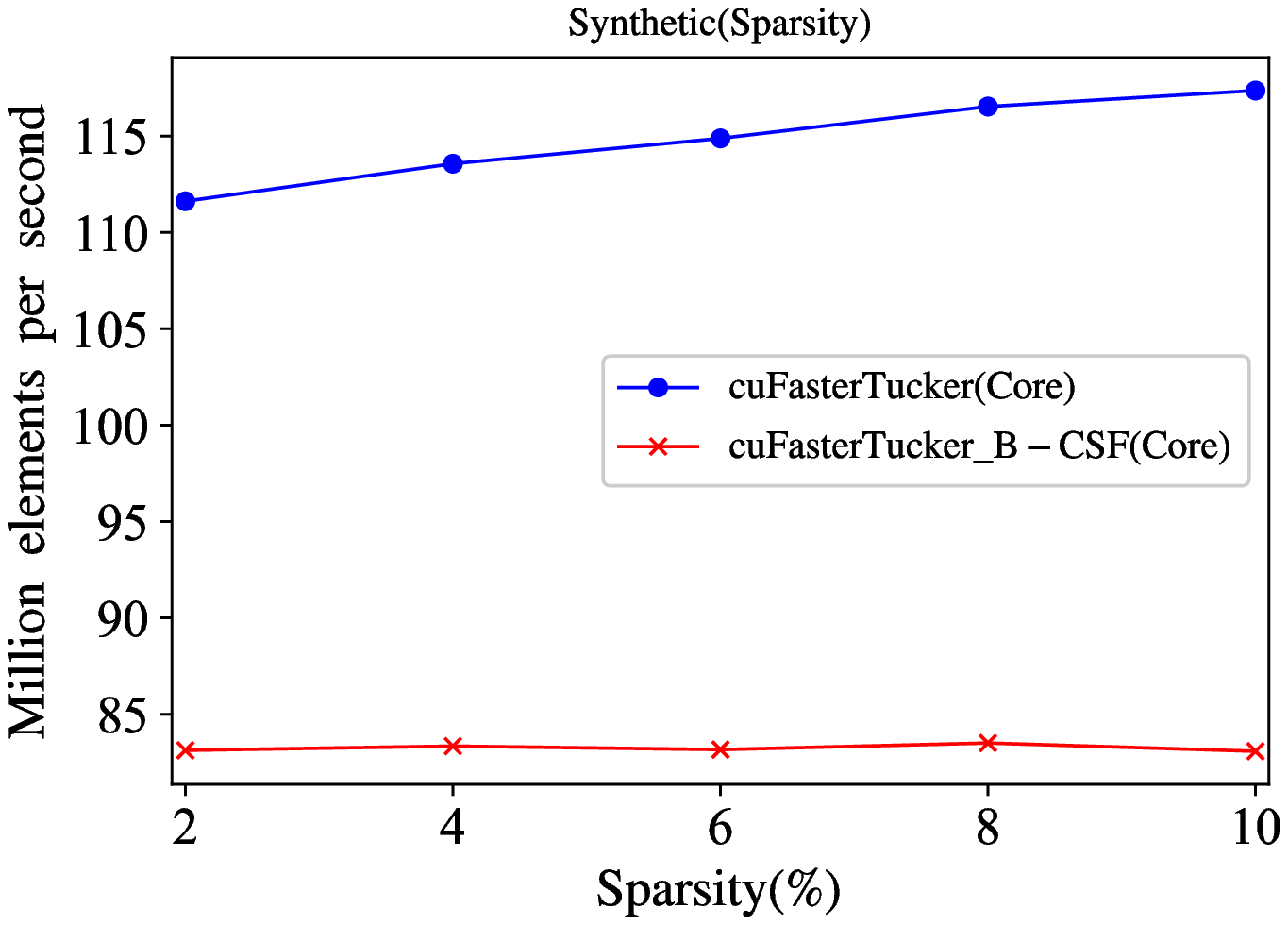}}
	\caption{(a) Adaptability of high-order tensors; (b)-(c) Adaptability of tensor sparsity.}
	\label{synthetic}
\end{figure*}

\subsection{Real-world accuracy}
We evaluate the accuracy of cuFasterTucker and its contrasting algorithms on the real-world datasets. 
The evaluation metrics are test Root Mean Square Error (RMSE) and test Mean Absolute Error(MAE) , 
which are widely used by recommender systems.
We randomly generate factor matrices and core matrices, which follow an average distribution. 
Each algorithm iterates $50$ times and records the test RMSE and test MAE after each iteration.
Figure \ref{convergence} shows the convergence curves of cuFasterTucker and its contrasting algorithms. 
Figures \ref{convergence(a)} and \ref{convergence(b)} describe the changes of RMSE and MAE on the Netflix dataset, respectively, 
and Figures \ref{convergence(c)} and \ref{convergence(d)} describe the changes of rmse and mae on the Yahoo!Music dataset, respectively.
It can be seen from Figure \ref{convergence} that the above algorithms all converge within acceptable accuracy.
The convergence curves of cuFasterTucker\_COO and cuFastTucker almost coincide, which is predictable, because the difference between the two is only reflected in the amount of calculation and other update methods are consistent.
cuFasterTucker converges a bit faster due to the data locality sensitivity of the B-CSF tensor storage format.

\subsection{Adaptability of high-order tensors}
We use synthetic datasets Synthetic(Order) to examine the adaptability of cuFasterTucker and its contrasting algorithms for high-order tensors. The orders of these tensors range from $3$ to $10$, the length of each order is fixed at $1000$,
and the number of non-zero elements is fixed at $100$ million.
We set $J_n=32, n \in \{N\}$ and $R=32$ for all algorithms and recorded the average time(seconds) for a single iteration. 
Figure \ref{synthetic(a)} shows the growth trend of the average single iteration time of these algorithms as the dimensionality increases.
Both cuFasterTucker\_COO and cuFasterTucker grow more slowly than cuFastTucker, depending on the computational complexity of the first two is much smaller than cuFastTucker.
It can be seen that cuFasterTucker is more suitable for high-order FastTucker decomposition than cuFastTucker.

\subsection{Adaptability of tensor sparsity} \label{Section Adaptability of tensor sparsity}
We use synthetic datasets Synthetic(Sparsity) to examine the adaptability of cuFasterTucker and its contrasting algorithms to tensors of varying sparsity.
The non-zero elements of these tensors range from $20$ million to $100$ million in increments of $20$ million, 
the order is fixed at $3$, and the length of each order is fixed at $1000$.
That is, five tensors with a sparsity of 2\% to 10\% whose growth is 2\%.
Since the number of non-zero elements in each tensor is not the same, 
it is not appropriate for us to use the average time of a single iteration for comparison.
We use the number of non-zero elements processed per second for comparison here.
Figures \ref{synthetic(b)} and \ref{synthetic(c)} show the efficiency of cuFasterTucker and its contrasting algorithms for different sparsity in updating factor and core matrices, respectively.
It can be seen that the efficiency of cuFasterTucker increases significantly with the increase of sparsity, 
while the efficiency of cuFasterTucker\_B-CSF does not change significantly.
As the data size increases, the amount of computation increases accordingly, 
and the sparsity increases, but the shared intermediate variables that cuFasterTucker needs to compute increases only slightly or even at all.
The results show that cuFasterTucker is more friendly to tensors with high sparsity, 
but its performance on tensors with low sparsity is not worse than other algorithms.

\section{Conclusion} \label{Section Conclusion}
We propose the FasterTucker algorithm, which has lower computational complexity than FastTucker and is more suitable for processing HOHDST. 
And we present a fine-grained parallel FasterTucker algorithm cuFasterTucker on GPU.
cuFasterTucker uses the SGD method to sequentially update the factor matrix or core matrix of each order to achieve convergence.
And cuFasterTucker uses the B-CSF tensor storage format to greatly improve the read and write efficiency during the update process while ensuring load balancing.
cuFasterTucker is about $14$ times and $6$ times faster than existing state-of-the-art algorithms in updating factor matrices and core matrices, respectively.
And the convergence speed is slightly improved.
More importantly, compared with the current state-of-the-art algorithms, cuFasterTucker performs well on high-order tensors, 
and the single iteration time is much smaller than the former, 
and the higher the order, the more obvious the gap.
Finally, cuFasterTucker performs better on denser tensors, 
a property that state-of-the-art algorithms do not have.
In future work, 
we will explore the FastTucker algorithm with faster convergence speed and extend it to new parallel hardware platforms as well as distributed platforms.

\bibliographystyle{IEEEtran}
\bibliography{bib}

\end{document}